\documentclass{eitnlp}

\usepackage{enumitem}
\usepackage[table]{xcolor}
\usepackage{longtable}
\usepackage{subcaption}
\usepackage{mdframed}
\usepackage{listings}
\usepackage{makecell}
\usepackage{multirow}
\usepackage{array}
\usepackage{siunitx}
\usepackage{fontawesome5}
\definecolor{darkblue}{rgb}{0, 0, 0.5}

\title{MMEB-V3: Measuring the Performance Gaps of Omni-Modality Embedding Models}
\runningtitle{MMEB-V3}

\author{%
\textbf{Haohang Huang\textsuperscript{1,*}}\;
\textbf{Xuan Lu\textsuperscript{1,2,*}}\;
\textbf{Mingyi Su\textsuperscript{4}}\;
\textbf{Xuan Zhang\textsuperscript{5}},\;
\textbf{Ziyan Jiang\textsuperscript{6}}\;\\
\textbf{Ping Nie\textsuperscript{4}},\;
\textbf{Kai Zou\textsuperscript{7}},\;
\textbf{Tomas Pfister\textsuperscript{3}},\;
\textbf{Wenhu Chen\textsuperscript{4}},\;
\textbf{Wei Zhang\textsuperscript{1}},\;\\
\textbf{Xiaoyu Shen\textsuperscript{1,\textdagger}},
\textbf{Rui Meng\textsuperscript{3}}\;
}
\affiliation{%
\textsuperscript{1}Eastern Institute of Technology, Ningbo\qquad
\textsuperscript{2}Shanghai Jiao Tong University\\
\textsuperscript{3}Google AI Research
\textsuperscript{4}University of Waterloo
\textsuperscript{5}NUS
\textsuperscript{6}UCSB
\textsuperscript{7}Netmind.ai
}
\authornotes{\textsuperscript{*}Equal contribution.\qquad
             \textsuperscript{\textdagger}Corresponding author.}
\contact{\texttt{lux1997@sjtu.edu.cn}\quad
         \texttt{haohang623@gmail.com}\quad
         \texttt{xyshen@eitech.edu.cn}}
\paperlinks{Published at COLM 2026 \\
\begin{center}

\vspace{-0.8em}

\href{https://github.com/TIGER-AI-Lab/VLM2Vec}
{\faGithub\; Code: VLM2Vec}
\href{https://huggingface.co/spaces/TIGER-Lab/MMEB-Leaderboard}
{\includegraphics[height=1.2em]{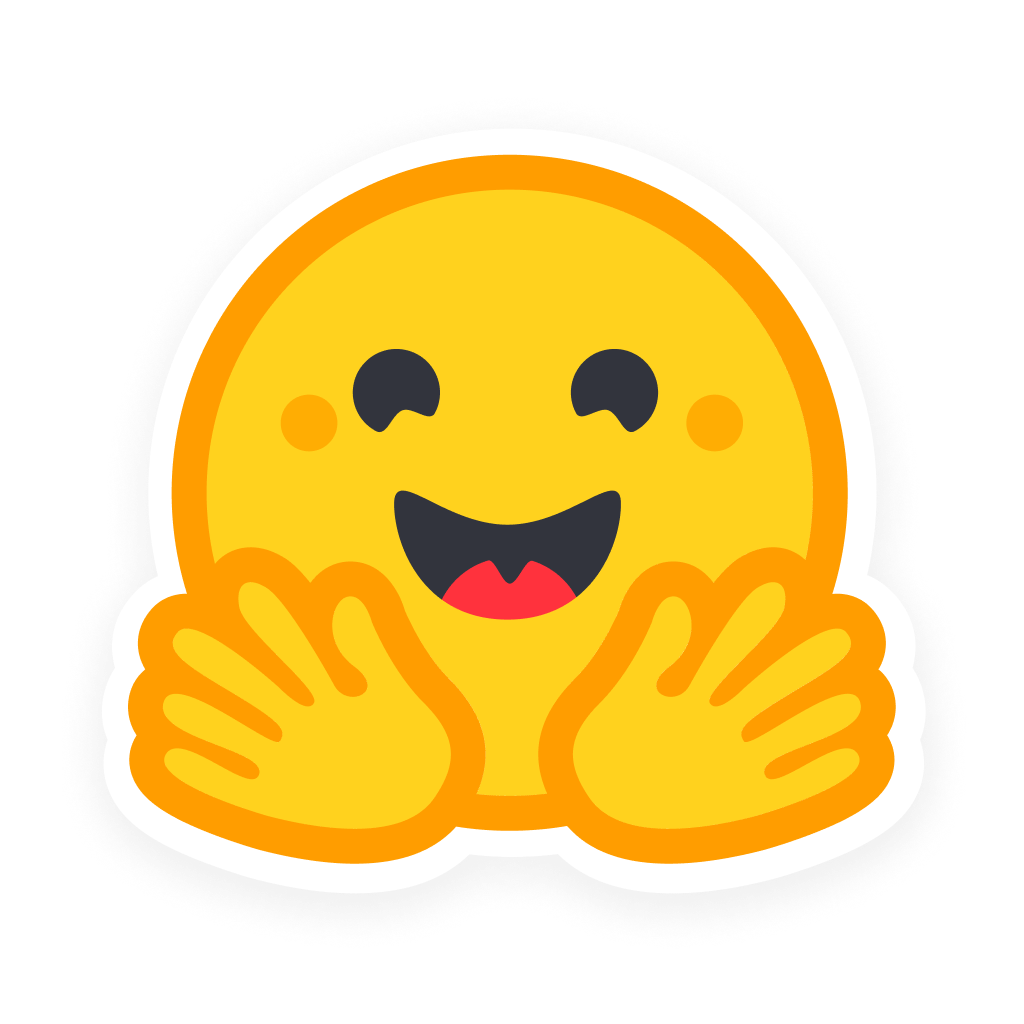}\; Leaderboard: MMEB-V3}
\end{center}}

\begin{document}

\maketitle

\begin{abstract}
Multimodal embedding models aim to map heterogeneous inputs, such as text, images, videos, and audio, into a shared semantic space. However, existing methods and benchmarks remain largely limited to partial modality coverage, making it difficult to systematically evaluate full-modality representation learning. In this work, we take a step toward the \textit{full-modality} setting. We introduce \textbf{MMEB-V3}, a comprehensive benchmark that evaluates embeddings across text, image, video, audio, as well as agent-centric scenarios. To enable more fine-grained diagnosis, we further construct \textbf{OmniSET (Omni-modality Semantic Equivalence Tuples)}, where semantically equivalent instances are represented across modalities, allowing us to disentangle semantic similarity from modality effects. Through experiments on MMEB-V3, we conduct a systematic analysis of full-modality embeddings and identify three key findings: (1) models often fail to retrieve the intended target modality; (2) cross-modal retrieval is highly asymmetric and dominated by query-modality bias; and (3) instruction-induced shifts are either insufficient or misaligned with the target modality, and therefore do not reliably improve retrieval. These results indicate that current multimodal embeddings are not yet capable of reliably enforcing modality constraints specified by instructions, and consequently fail to exhibit consistent modality-aware retrieval behavior. We hope MMEB-V3 provides a useful benchmark for understanding and diagnosing these limitations, and for guiding future research on full-modality embeddings.
\end{abstract}

\section{Introduction}
Multimodal embeddings \citep{zhang2025gme, meng2025vlm2vecv2advancingmultimodalembedding} are foundational to modern machine learning, mapping heterogeneous inputs---such as text, images, videos, and audio---into a unified, fixed-dimensional vector space. These representations power a vast ecosystem of applications, from semantic retrieval and recommendation systems to complex decision-making pipelines. As the field evolves, research has shifted from isolated unimodal encoders \citep{clip} toward \emph{unified multimodal embedding models} that align diverse modalities. Such unified spaces are increasingly critical for emerging paradigms like multimodal Retrieval-Augmented Generation (RAG) and intelligent AI agents, which must retrieve and act upon information seamlessly across different sensory inputs \cite{huang2026search}.

Despite this progress, we identify a fundamental but underexplored challenge: \textbf{modality as an explicit instruction constraint}. Existing frameworks implicitly assume that modality is either perfectly aligned with semantics or acts as a passive attribute. In practice, however, modality often serves as an explicit requirement specified by a user or system. For example, a user may issue queries such as "\textit{find an audio clip of a cat meowing}" or "\textit{retrieve a video showing a cat jumping}", where modality is a mandatory constraint rather than an optional attribute. 
This reveals a key limitation: \textit{current embedding models often fail to interpret modality as an explicit instruction constraint, instead treating it as a byproduct of semantic similarity}, leading to semantically relevant but modality-mismatched retrieval results. 
This issue is particularly acute in agent-centric settings, where retrieval is a prerequisite for downstream actions such as tool invocation, GUI interaction, or memory access. Errors in modality understanding---such as retrieving semantically relevant but modality-mismatched results---can lead to catastrophic failures in the decision-making loop. 
However, current benchmarks such as MMEB~\citep{meng2025vlm2vecv2advancingmultimodalembedding} and UMR~\citep{zhang2025gme} provide limited support for evaluating this problem. They focus primarily on cross-modal alignment (e.g., text-to-image) and lack systematic evaluation of (1) full-modality coverage (especially audio and visual documents) and (2) instruction-following behavior in complex retrieval scenarios.

To systematically \textbf{measure and diagnose limitations in modality-aware, instruction-conditioned embedding behavior}, we introduce \textbf{MMEB-V3}, a comprehensive full-modality benchmark designed to evaluate embeddings under realistic, instruction-conditioned settings. MMEB-V3 extends prior work along four key dimensions: 
(1) \textbf{Audio Tasks}, expanding coverage to include audio classification, temporal grounding and cross-modal retrieval; 
(2) \textbf{Text Tasks}, incorporating instruction-following, reasoning, and multi-condition constraints; 
(3) \textbf{Agent Tasks}, evaluating embeddings in agent-centric scenarios such as tool retrieval, memory retrieval, and GUI control; and 
(4) \textbf{OmniSET (Omni-modality Semantic Equivalence Tuples)}, which organize semantically equivalent content across modalities into unified tuples, enabling controlled analysis that disentangles semantic content from modality effects. Through experiments on MMEB-V3, we conduct a systematic analysis of embedding behavior under instruction-conditioned settings. Our results show that while existing models perform well on standard semantic alignment, they struggle to enforce modality constraints specified by instructions. In many cases, retrieval is dominated by the original query modality rather than the instructed target modality, leading to systematic modality mismatches. These findings highlight a key limitation in current multimodal embeddings: modality constraints are not reliably enforced as part of the instruction.

Our main contributions are as follows:
\begin{itemize}
\item We introduce \textbf{MMEB-V3}, a comprehensive benchmark for evaluating full-modality embeddings under instruction-conditioned settings across diverse tasks.
\item We provide a \textbf{systematic diagnostic analysis} of modality-aware behavior, enabled by \textbf{OmniSET (Omni-modality Semantic Equivalence Tuples)}, a controlled design that disentangles semantic content from modality effects.
\item We conduct a \textbf{comprehensive analysis of modality-constrained retrieval}, showing that (i) models often fail to retrieve the intended target modality, (ii) retrieval is asymmetric and biased toward the query modality, and (iii) instruction-induced shifts are insufficient or misaligned with the target modality.
\end{itemize}

\section{Related Work}
\paragraph{Multimodal Embedding Benchmarks.}
The evolution of multimodal embedding benchmarks reflects a transition from coarse-grained cross-modal alignment to fine-grained, task-diverse evaluation. Early efforts primarily focused on image--text alignment, with benchmarks such as MSCOCO~\citep{MSCOCO}, Flickr30K~\citep{plummer2015flickr30k} and UMR~\citep{zhang2025gme} establishing standard protocols for retrieval and matching. While foundational, these benchmarks are largely constrained to static visual inputs and brief textual descriptions. Subsequent research has expanded this scope to specialized domains: ViDoRe-v2~\citep{macé2025vidorebenchmarkv2raising} introduces document-level visual retrieval, while QVHighlights~\citep{lei2021qvhighlightsdetectingmomentshighlights} and FineVideo~\citep{Farré2024FineVideo} address temporal dynamics in video retrieval.
Despite these advancements, a significant gap persists in evaluating instruction-driven, full-modality generalizability. While MTEB~\citep{muennighoff2023mtebmassivetextembedding} provides a comprehensive suite for unimodal text embeddings, it lacks multimodal support; conversely, MMEB-V2~\citep{meng2025vlm2vecv2advancingmultimodalembedding} and M-BEIR~\citep{wei2023uniir} integrate images, videos, and documents but omit other critical modalities such as audio. Furthermore, emerging agentic tasks—including tool retrieval~\citep{lu2025toolsunderdocumentedsimpledocument}, GUI control~\citep{gui} and memory retrieval~\citep{zhao2026lmeblonghorizonmemoryembedding}—increasingly rely on embedding-based solutions, yet they remain excluded from existing comprehensive benchmarks. Existing multimodal embedding benchmarks mainly focus on semantic alignment across modalities (e.g., text–image retrieval), while more complex scenarios—such as instruction-conditioned retrieval, multi-condition reasoning, and agent tasks (e.g., tool, memory, and GUI retrieval)—remain underexplored.

\paragraph{Multimodal Embedding Models.}
Multimodal embedding models aim to learn shared representations for heterogeneous modalities (e.g., text, images, audio, and video) within a unified framework. 
Early approaches typically follow a dual-encoder paradigm, aligning modality-specific encoders in a common space, as exemplified by CLIP~\citep{clip}, ALIGN~\citep{align}, and their extensions to video and audio.
Building on this foundation, recent work explores partial modality unification for embedding-based retrieval. 
Models such as GME~\citep{zhang2025gme}, MM-Embed~\citep{lin2025mmembeduniversalmultimodalretrieval}, VLM2Vec~\citep{jiang2025vlm2vectrainingvisionlanguagemodels}, and VLM2Vec-V2~\citep{meng2025vlm2vecv2advancingmultimodalembedding} unify subsets of modalities and achieve strong performance on their target benchmarks, often leveraging MLLM backbones or contrastive objectives.
More recent efforts move toward broader modality coverage. 
Omni-Embed-Nemotron~\citep{xu2025omniembednemotronunifiedmultimodalretrieval} supports retrieval across text, image, audio, and video, while OmniRet~\citep{huynh2026omniretefficienthighfidelityomni} improves efficiency via multimodal token resampling and pooling strategies. 
WAVE~\citep{tang2025wavelearningunified} focuses on unified audio–visual embeddings, enabling cross-modal retrieval between audio and video.
To complement these developments, we introduce MMEB-V3, which extends prior benchmarks with audio modalities, complex text retrieval, and agent tasks, enabling more comprehensive evaluation under realistic, instruction-conditioned settings.


\section{MMEB-V3: A Unified Evaluation Framework for Omni-Modality Embeddings}

\subsection{Modality Coverage and Task Diversity}

\textbf{MMEB-V3} extends MMEB-V2 by adding \textbf{111} new tasks (see Table~\ref{tab:benchmark_overview}). It provides an evaluation framework covering four modalities—text, image, video, and audio—encompassing a total of \textbf{190} tasks. Beyond expanding modality coverage, MMEB-V3 organizes tasks across diverse cross-modal directions (e.g., A2I, T2A, A2V), allowing evaluation under a range of modality interactions. On top of this modality space, it includes multiple task types, such as audio classification, cross-modal retrieval, multi-condition text retrieval, and agent tasks. This design reflects both modality diversity and task diversity, with tasks evaluated under specific modality configurations rather than in isolation. 

\begin{figure}[t]
    \centering
    \includegraphics[width=0.99\linewidth]{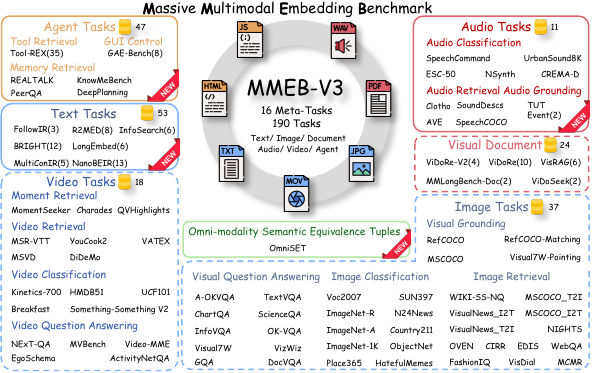}
    \caption{MMEB-V3 Overview: New Additions—Agent Tasks, Complex Text Retrieval, Audio Tasks and Equivalence Tuples; Built upon Image, Video, and VisDoc from MMEB-V2}
    \label{fig:mmeb-v3}
\end{figure}

\begin{figure*}[th]
    \centering
    \begin{subfigure}{0.6\textwidth}
        \centering
        \includegraphics[width=\linewidth]{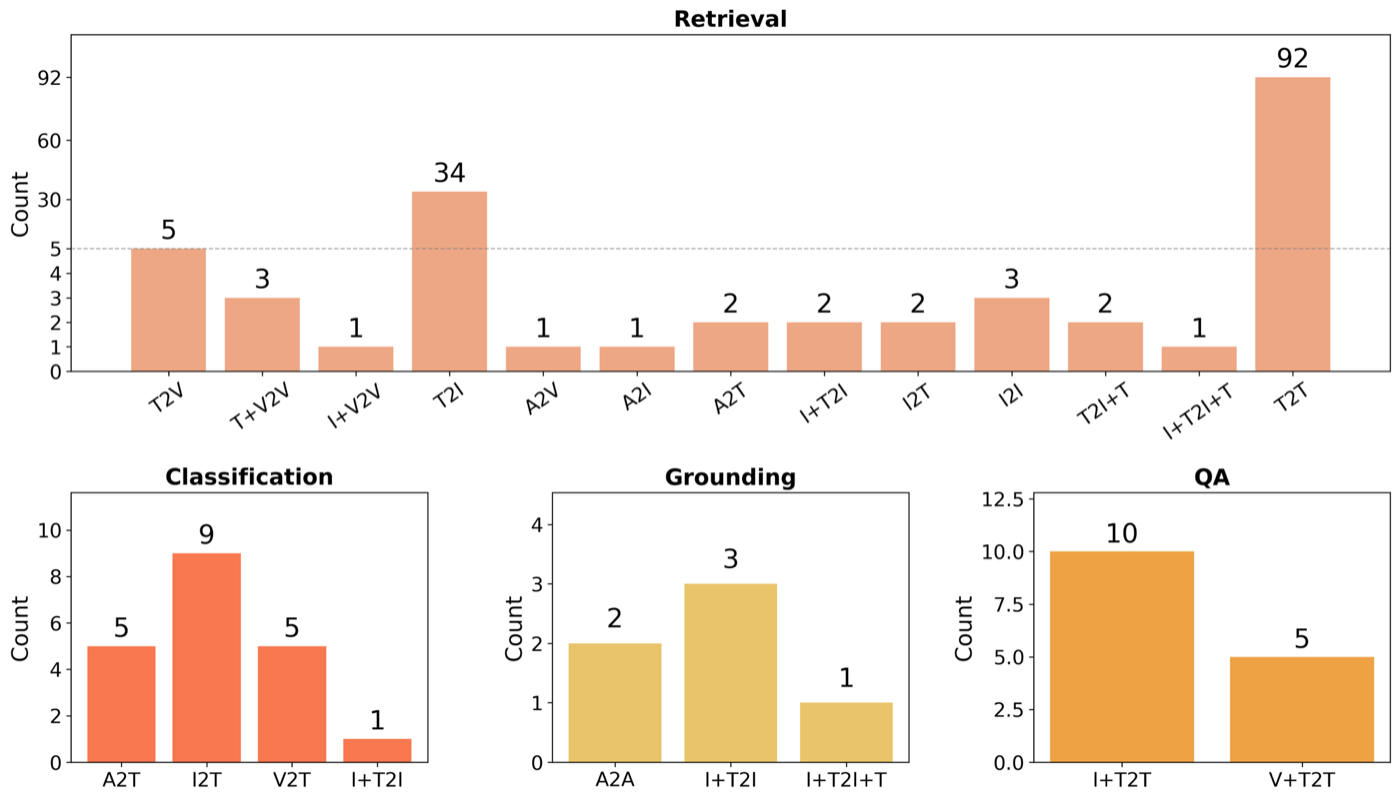}
        \caption{Distribution of modality combinations across task types in MMEB-V3. This illustrates the comprehensive modality coverage and the diversity of cross-modal interactions and task types.}
        \label{fig:modality_diversity}
    \end{subfigure}
    \hfill
    \begin{subfigure}{0.38\textwidth}
        \centering
        \includegraphics[width=\linewidth]{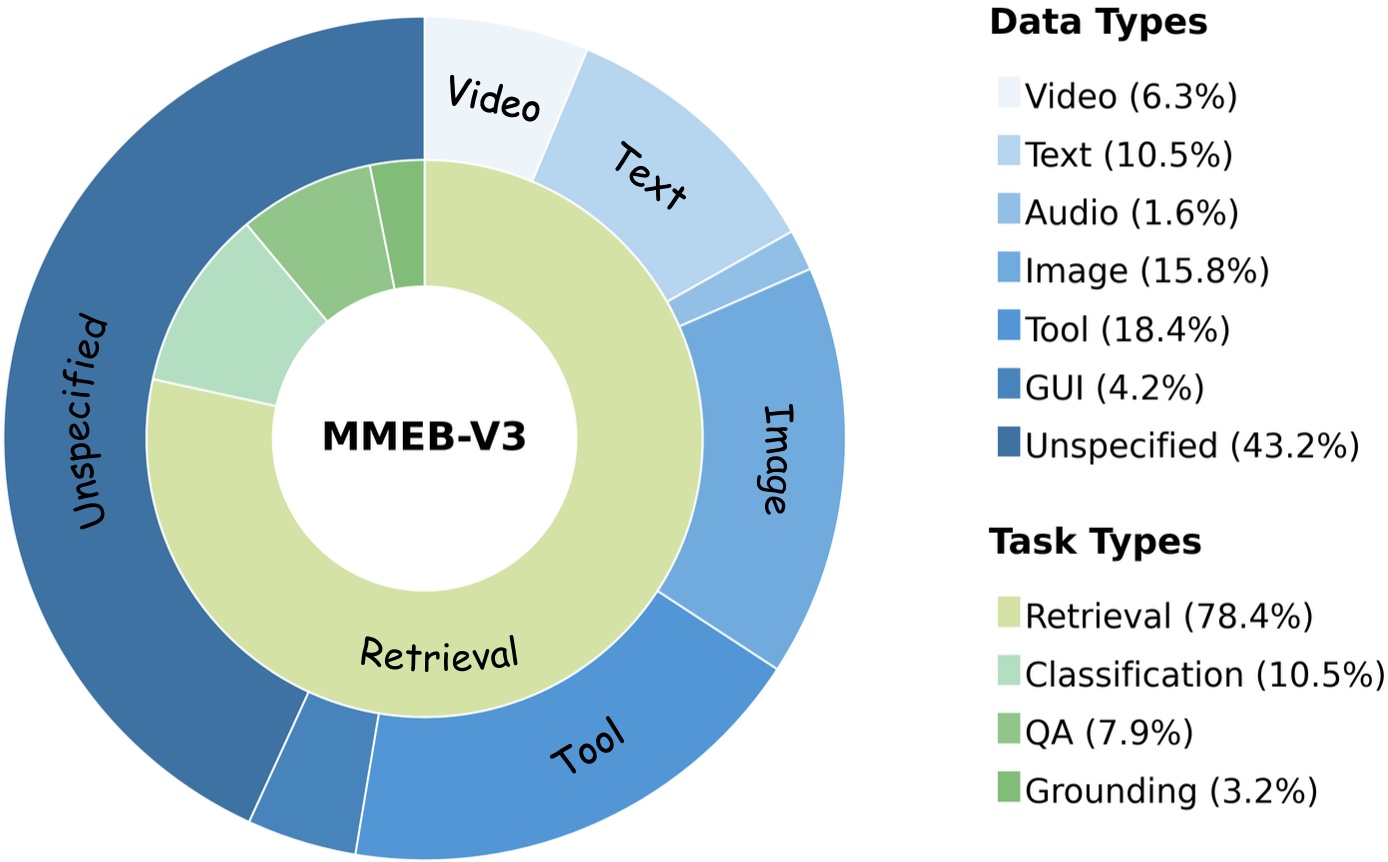}
        \caption{Distribution of instruction patterns, including data-type constraints and task-type diversity. This highlights that instructions vary in their data requirements as well as the diversity of underlying task types.}
        \label{fig:instruction_diversity}
    \end{subfigure}
    \caption{ Diversity of modalities, tasks, and instruction patterns in MMEB-V3\textbf{ (190 tasks)}.
(a) Distribution of modality combinations across task types, demonstrating comprehensive modality coverage and rich cross-modal interactions across diverse tasks. 
(b) Distribution of instruction patterns, including data-type constraints and task-type diversity, showing that instructions are not homogeneous but vary in both data requirements and associated task contexts.}
    \label{fig:diversity}
\end{figure*}

In addition to standard evaluation, MMEB-V3 includes a diagnostic component, OmniSET (Omni-modality Semantic Equivalence Tuples), which groups semantically equivalent instances across modalities to facilitate controlled comparisons. OmniSET is not used for leaderboard evaluation, but provides a setting to analyze the relationship between semantic content and modality effects. Figure~\ref{fig:mmeb-v3} provides an overview of MMEB-V3. Figure~\ref{fig:modality_diversity} further illustrates the distribution of tasks across modalities and task categories.

\paragraph{Audio Tasks.}
MMEB-V3 incorporates diverse audio tasks, including classification (e.g., ESC-50~\citep{esc50}, UrbanSound8K~\citep{urbansound}, NSynth~\citep{nsynth2017}), cross-modal retrieval (e.g., Clotho~\citep{2020clotho}, SoundDescs~\citep{sounddescs}, AVE~\citep{ave}, SpeechCOCO~\citep{speechcoco}), and temporal grounding (TUT Sound Events~\citep{tutsound2017}). For TUT Sound Events, we construct two tasks with different difficulty levels and use their average score as the final temporal grounding metric. The audio tasks evaluate both intra-modal understanding and alignment between audio and other modalities, providing a basis to assess acoustic-semantic representations.

\paragraph{Text Tasks.}
Beyond standard semantic retrieval, MMEB-V3 includes text scenarios that involve instruction following (FollowIR~\citep{weller-etal-2025-followir}, InfoSearch~\citep{zhou2025beyond}), reasoning (BRIGHT~\citep{su2025bright}, R2MD~\citep{li2025r2medbenchmarkreasoningdrivenmedical}), long-context understanding (LongEmb~\citep{zhu-etal-2024-longembed}), and multi-condition matching (MultiConIR~\citep{lu-etal-2025-multiconir}). A compact general retrieval benchmark (nanoBEIR~\citep{thakur2021beirheterogenousbenchmarkzeroshot}) is also included to provide coverage of general retrieval settings. These datasets are composed of multiple tasks, resulting in a total of 53 tasks. The final Text score is computed as the average of NDCG@5 across these tasks.

\paragraph{Agent Tasks.}
MMEB-V3 includes agent tasks, such as tool retrieval (Tool-REX~\citep{lu2025toolsunderdocumentedsimpledocument}), GUI trajectory retrieval (GAE-Bench~\citep{gui}), and memory retrieval (KnowMeBench~\citep{wu2026knowmebenchbenchmarkingpersonunderstanding}, REALTALK~\citep{lee2025realtalk21dayrealworlddataset}, PeerQA~\citep{baumgärtner2025peerqascientificquestionanswering}, and DeepPlanning~\citep{zhang2026deepplanningbenchmarkinglonghorizonagentic}). These tasks involve selecting tools or actions, or retrieving relevant information from past memory based on multimodal inputs and structured instructions. These datasets are composed of multiple tasks, resulting in a total of 47 tasks. The final Agent score is computed as the average of Hit@1 across these tasks.

\paragraph{Cross-Modal Equivalence Modeling.}
We include a diagnostic component, OmniSET (Omni-modality Semantic Equivalence Tuples), which groups semantically equivalent instances across modalities into tuples $\{x^T, x^I, x^V, x^A\}$. OmniSET is not used for leaderboard evaluation, but provides a setting for controlled comparisons between semantic content and modality effects. It enables analysis of how models handle modality as an explicit constraint under instruction-conditioned retrieval.

\subsection{Instruction Diversity across Data Types and Tasks}

We describe instruction diversity in MMEB-V3 from two perspectives: \textbf{target-type constraints} and \textbf{task-type diversity}. First, \textbf{Data-type constraints} refer to the expected type of returned data, including modalities such as \textbf{text}, \textbf{image}, \textbf{audio}, and \textbf{video}, as well as structured data such as \textbf{tools} and \textbf{GUI elements}. This dimension captures whether and how instructions constrain the modality or representation of the data. Second, \textbf{Task-type diversity} corresponds to the range of operations associated with different tasks. In MMEB-V3, these mainly include \textbf{classification}, \textbf{retrieval}, \textbf{grounding}, and \textbf{QA}, covering behaviors ranging from label prediction and candidate selection to localization and question answering.

Together, these perspectives offer a way to describe instruction diversity in MMEB-V3. This formulation shows that tasks vary in both \emph{what} type of output is required and \emph{what} operation is involved. Figure~\ref{fig:instruction_diversity} visualizes the distribution of these patterns across MMEB-V3.

\begin{table*}[t]
\centering
\small
\setlength{\tabcolsep}{3pt}
\renewcommand{\arraystretch}{1.2}
\resizebox{\textwidth}{!}{%
\begin{tabular}{c c c c c c c}
\toprule
\textbf{Task} & \textbf{Task type} & \textbf{Query MOD} & \textbf{Target MOD} & \textbf{Domain} & \textbf{\#Query} & \textbf{\#Candidates} \\
\midrule

\multicolumn{7}{c}{\textbf{Audio Tasks (11 tasks)}} \\
\midrule

ESC-50           & Audio Classification     & A & T &Environmental sounds  &2K  &50  \\
\rowcolor{gray!16}
UrbanSound8K     & Audio Classification     & A & T &Environmental sounds  &1.7K  &10  \\
NSynth           & Audio Classification     & A & T &Music  &1K  &10  \\
\rowcolor{gray!16}
Speech Commands  & Audio Classification     & A & T &Human Voice  &1K  &36  \\
CREMAD           & Audio Classification     & A & T &Human Voice  &7.4K  &6  \\
\rowcolor{gray!16}
Clotho           & Audio Retrieval          & T & A &Open  &5.2K  &1K  \\
Sound-Descs      & Audio Retrieval          & T & A &Open  &1K  &4.9K  \\
\rowcolor{gray!16}
AVE              & Audio Retrieval          & A & V &Acoustic Events  &402  &402  \\
SpeechCOCO       & Audio Retrieval          & A & I &Open  &1K  &10K  \\
\rowcolor{gray!16}
TUT Sound Events (2) & Audio Temporal Grounding  & A  & A &Acoustic Events  &659  &$51 \sim 106$  \\

\midrule
\multicolumn{7}{c}{\textbf{Text Tasks (53 tasks)}} \\
\midrule

FollowIR (3)         & Instruction-following  & T & T & Open &104  &98K  \\
\rowcolor{gray!16}
InfoSearch (6)       & Instruction-following  & T & T & Open &1.6K  &6.3K  \\

BRIGHT (12)           & Reasoning Retrieval      & T & T & Open &1.4K  &1.3M  \\
\rowcolor{gray!16}
R2MD (8)             & Reasoning Retrieval      & T & T & Medical &876  &357K \\

LongEmb (6)          & Long-context Retrieval  & T & T & Open &13K  &2.7K  \\
\rowcolor{gray!16}
MultiConIR (5)       & Multi-condition Retrieval & T & T & Open &9K  &25K  \\
nanoBEIR (13)         & General Text Retrieval  & T & T & Open &649  &56K  \\

\midrule
\multicolumn{7}{c}{\textbf{Agent Tasks (47 tasks)}} \\
\midrule

Tool-REX (35)         & Tool Retrieval  & T & T & Tool Use &7.9K  &44K  \\
\rowcolor{gray!16}
GAE-Bench (8)        & GUI Control & I/T & I/T & GUI &8.2K  &78K  \\
KnowMeBench        & Agent Memory Retrieval & T & T & Episodic Memory &2K  &27K  \\
\rowcolor{gray!16}
REALTALK        & Agent Memory Retrieval & T & T & Dialogue Memory &679  &8.9K  \\

PeerQA        & Agent Memory Retrieval & T & T & Semantic Memory &136  &18.6K  \\
\rowcolor{gray!16}
DeepPlanning        & Agent Memory Retrieval & T & T & Procedural Memory &120  &19.8K  \\

\midrule
\multicolumn{7}{c}{\textbf{Omni-modality Semantic Equivalence Tuples}} \\
\midrule

OmniSET  & Omni-modality Semantic Equivalence Tuples & T/I/V/A & T/I/V/A & Open &1.2K &8.2K \\

\bottomrule
\end{tabular}%
}
\caption{The statistics of MMEB-V3 are summarized below. Compared to MMEB-V2, MMEB-V3 introduces \textbf{111 new tasks} across three major categories: audio tasks, agent tasks, and complex text retrieval tasks, resulting in a total of \textbf{190 tasks}. We report the detailed statistics of these newly added tasks. We also include statistics of OmniSET as a diagnostic component for controlled analysis. OmniSET is used for comparative analysis of modality effects and is not part of the leaderboard evaluation. Modalities (MOD): T (Text), I (Image), V (Video), and A (Audio).}
\label{tab:benchmark_overview}
\end{table*}

\section{Experiments}

\subsection{Experimental Settings}

\paragraph{Metrics.} MMEB-V3 uses task-appropriate metrics. We adopt \textbf{Hit@1} for audio, image, video, and agent tasks, where the number of relevant targets is typically small and the evaluation focuses on identifying a highly relevant match. We use \textbf{NDCG@5} for text and VisDoc tasks, which involve multiple relevant candidates and require fine-grained ranking. This design aligns evaluation metrics with task characteristics.
\paragraph{Baselines.}
We evaluate against two groups of baselines, including omni-modal embedding models (Omni-Embed-Nemotron~\citep{xu2025omniembednemotronunifiedmultimodalretrieval} and WAVE~\citep{tang2025wavelearningunified}) and vision-language embedding models (Qwen3-VL-Embedding~\citep{qwen3vlembedding}, VLM2Vec-V2.0~\citep{meng2025vlm2vecv2advancingmultimodalembedding}, VLM2Vec~\citep{jiang2024vlm2vec}, E5-Omni~\citep{chen2026e5}, LCO-Embedding-Omni~\citep{LCO} and GME~\citep{zhang2025gme}).

\subsection{Main Results}

Table~\ref{tab:main_results_overall_modalities} presents a comparison of representative multimodal embedding models on MMEB-V3, covering text, image, video, audio, visual document, and agent tasks. Full results across all tasks are provided in Appendix~\ref{sec: detailed scores}. A general pattern can be observed: no single model achieves optimal performance across all task categories. Models that perform well on certain modalities or tasks often exhibit weaker performance on others, suggesting trade-offs in current embedding approaches. 

For example, models such as Qwen3-VL-Embedding achieve strong performance on text, image, and video tasks, but lack native audio capability, which may limit their applicability in settings involving audio. 
Fully multimodal models (e.g., E5-Omni) show relatively stable performance across modalities, yet do not consistently outperform specialized models within specific task categories. To further understand these trade-offs, Table~\ref{tab:main_results_audio_text_agent} provides a fine-grained breakdown of newly introduced audio, text, and agent tasks. We observe that several task types are relatively challenging across models, including audio retrieval, as well as complex text and agent tasks involving reasoning, multi-condition constraints, and long-context understanding. This may indicate limitations in capturing fine-grained semantic structure and implicit constraints.

Overall, these findings suggest that the observed trade-offs are associated with both modality differences and task complexity, as well as instruction requirements, pointing to the importance of developing more unified and instruction-aware embedding models.

\begin{table*}[t]
\centering
\small
\setlength{\tabcolsep}{4pt}
\renewcommand{\arraystretch}{1.2}

\resizebox{\textwidth}{!}{%
\begin{tabular}{lcccccccccccccccc}
\toprule
\multirow{2}{*}{\textbf{Model}} 
& \multicolumn{4}{c}{\textbf{Audio}}
& \multicolumn{6}{c}{\textbf{Text}}
& \multicolumn{4}{c}{\textbf{Agent}}
& \multirow{2}{*}{\textbf{All$^{*}$}}
& \multirow{2}{*}{\textbf{All}} \\
\cmidrule(lr){2-5}
\cmidrule(lr){6-11}
\cmidrule(lr){12-15}
& \textbf{CLS} & \textbf{RET} & \textbf{TG} & \textbf{Overall}
& \textbf{RR} & \textbf{IF} & \textbf{LC} & \textbf{MC} & \textbf{GR} & \textbf{Overall}
& \textbf{Tool} & \textbf{GUI} & \textbf{Memory} & \textbf{Overall}
& & \\
\midrule

\textbf{\# of Datasets $\rightarrow$}
& 5 & 4 & 2 & 11
& 20 & 9 & 6 & 5 & 13 & 53
& 35 & 8 & 4 & 47
& 100 & 111 \\

\midrule

Qwen3-VL-Embedding (2B)
& {-} & {-} & {-} & {-}
& 16.6 & 46.0 & 53.1 & 61.2 & 58.0 & 40.1
& \textbf{42.6} & 30.4 & 28.4 & \textbf{39.3}
& 39.2 & 35.4 \\

\rowcolor{gray!16}
Qwen3-VL-Embedding (8B)
& {-} & {-} & {-} & {-}
& \textbf{20.0} & 49.5 & \textbf{58.0} & 61.2 & 61.2 & \textbf{43.3}
& 41.3 & 33.5 & 22.8 & 38.4
& \textbf{40.6} & 36.5 \\

VLM2Vec-Qwen2VL (7B)
& {-} & {-} & {-} & {-}
& 7.2 & 30.2 & 5.9 & 40.8 & 41.7 & 22.6
& 19.8 & 21.4 & 15.9 & 19.7
& 21.0 & 19.0 \\

\rowcolor{gray!16}
VLM2Vec-V2.0 (2B)
& {-} & {-} & {-} & {-}
& 9.1 & 32.8 & 11.5 & 50.3 & 41.2 & 25.1
& 27.6 & \textbf{36.2} & 23.3 & 28.7
& 26.5 & 23.9 \\

GME (7B)
& {-} & {-} & {-} & {-}
& 14.1 & \textbf{58.6} & 17.8 & 59.0 & \textbf{62.5} & 38.2
& 39.0 & 30.0 & 17.1 & 35.6
& 36.4 & 32.8 \\

\rowcolor{gray!16}
WAVE (7B)
& 52.3 & 12.9 & 21.5 & 31.8
& 3.6 & 34.0 & 2.6 & 22.9 & 18.6 & 14.1
& 11.9 & 11.8 & 5.7 & 11.3
& 14.3 & 14.3 \\

Omni-Embed-Nemotron (3B)
& 53.7 & 18.8 & 29.1 & 36.5
& 15.3 & 42.3 & 50.0 & \textbf{69.7} & 56.9 & 39.1
& 38.1 & 32.0 & \textbf{32.2} & 36.5
& 37.6 & \textbf{37.6} \\

\rowcolor{gray!16}
LCO-Embedding-Omni (7B)
& \textbf{64.3} & 19.9 & \textbf{36.8} & \textbf{43.2}
& 9.5 & 55.4 & 12.6 & 61.3 & 52.1 & 33.0
& 29.0 & 25.0 & 23.0 & 27.8
& 30.9 & 30.9 \\

E5-Omni (3B)
& 41.4 & 22.0 & 21.6 & 30.8
& 10.9 & 47.4 & 8.3 & 60.7 & 34.4 & 27.2
& 37.7 & 35.6 & 32.3 & 36.9
& 31.2 & 31.2 \\

\rowcolor{gray!16}
E5-Omni (7B)
& 63.0 & \textbf{23.5} & 32.2 & 43.0
& 12.0 & 44.8 & 9.7 & 56.9 & 35.9 & 27.4
& 37.4 & 38.0 & 27.3 & 36.7
& 32.6 & 32.6 \\

\bottomrule
\end{tabular}%
}

\caption{Performance comparison across Audio, Text, and Agent tasks.
CLS: classification; RET: retrieval; TG: temporal grounding; RR: reasoning retrieval; IF: instruction following; LC: long-context retrieval; MC: multi-condition retrieval; GR: general retrieval; Memory: agent memory retrieval.
Tool: tool retrieval; GUI: GUI control.
\textit{$^{*}$ indicates that the overall score is averaged over available tasks only; the final All column averages over all modalities, treating missing ones as 0.}}
\label{tab:main_results_audio_text_agent}
\end{table*}

\begin{table*}[t]
\centering
\small
\setlength{\tabcolsep}{4pt}
\renewcommand{\arraystretch}{1.2}
\resizebox{0.9\textwidth}{!}{%
\begin{tabular}{lcccccccc}
\toprule
\textbf{Model} & \textbf{Image} & \textbf{Video} & \textbf{VisDoc} & \textbf{Audio} & \textbf{Text} & \textbf{Agent} & \textbf{All$^{*}$} & \textbf{All} \\
\midrule

\textbf{\# of Datasets $\rightarrow$} 
& 37 & 18 & 24 & 11 & 53 & 47 & 179 & 190 \\
\midrule

Qwen3-VL-Embedding (2B)
&69.5 &55.9 &70.6 &{-} &40.1 &\textbf{39.3} &51.4 &48.4 \\

\rowcolor{gray!16}
Qwen3-VL-Embedding (8B)
&\textbf{72.1} &\textbf{58.6} &70.9 &{-} &\textbf{43.3}&38.4 &\textbf{53.0} &\textbf{49.9} \\

VLM2Vec-Qwen2VL (7B)
&63.6 &33.8 &32.6 &{-} &22.6 &19.7 &32.7 &30.8 \\

\rowcolor{gray!16}
VLM2Vec-V2.0 (2B)
&63.3 &34.7 &68.6 &{-} &25.1 &28.7 &40.6 &38.2 \\

GME (7B)
&55.2 &38.4 &75.2 &{-} &38.2 &35.6 &45.7 &43.0 \\

\rowcolor{gray!16}
WAVE (7B)
&41.5 &43.1 &42.8 &31.8 &14.1 &11.3 &26.3 &26.3 \\

Omni-Embed-Nemotron (3B) & 43.4 & 41.4 & 71.1 & 36.5 & 39.1 & 36.5 &43.5 &43.5 \\

\rowcolor{gray!16}
LCO-Embedding-Omni (7B) 
&49.6 &48.4 &68.7 &\textbf{43.2} &33.0 &27.8 &41.3 &41.3 \\

E5-Omni (3B) & 63.5 & 48.9 & 73.5 & 30.8 & 27.2 & 36.9 &44.6 &44.6 \\

\rowcolor{gray!16}
E5-Omni (7B) & 70.5 & 50.8 & \textbf{75.4} & 43.0 & 27.4 & 36.7 &47.1 &47.1 \\

\bottomrule
\end{tabular}%
}
\caption{
Performance comparison across Image, Video, VisDoc, Audio, Text, and Agent tasks.
\textit{$^{*}$ indicates that the overall score is averaged over available tasks only; the final All column averages over all modalities, treating missing ones as 0.}
}
\label{tab:main_results_overall_modalities}
\end{table*}

\section{Analysis}
\label{sec:analysis}

We analyze three representative embedding models on the OmniSET dataset: Omni-Embed-Nemotron-3B, WAVE, and Qwen3-VL-Embedding-8B. 
OmniSET provides semantically aligned instances across modalities, enabling controlled analysis of modality effects and instruction-conditioned retrieval.


\begin{table*}[t]
\centering
\setlength{\tabcolsep}{2pt}
\renewcommand{\arraystretch}{1.5}
\resizebox{\textwidth}{!}{%
\begin{tabular}{@{}c l cccccccccccc@{}}
\toprule
\textbf{Model} & \textbf{Metric} 
& \textbf{T2I} & \textbf{T2V} & \textbf{T2A} 
& \textbf{I2T} & \textbf{I2V} & \textbf{I2A} 
& \textbf{V2T} & \textbf{V2I} & \textbf{V2A} 
& \textbf{A2T} & \textbf{A2I} & \textbf{A2V} \\
\midrule

\multirow{3}{*}{\makecell[c]{Omni-Embed\\Nemotron}}
& Hit@1 
& 0.0 & 1.5 & 0.0 
& 0.0 & 100.0 & 0.0 
& 0.0 & 5.0 & 0.0 
& 100.0 & 0.0 & 0.0 \\

& \cellcolor{gray!12}MRR
& \cellcolor{gray!12}4.1 & \cellcolor{gray!12}16.4 & \cellcolor{gray!12}6.4
& \cellcolor{gray!12}10.4 & \cellcolor{gray!12}100.0 & \cellcolor{gray!12}1.7
& \cellcolor{gray!12}2.4 & \cellcolor{gray!12}19.4 & \cellcolor{gray!12}4.0
& \cellcolor{gray!12}100.0 & \cellcolor{gray!12}4.5 & \cellcolor{gray!12}29.8 \\

& Top-10 DM
& T(84.9\%) & T(84.7\%) & T(75.6\%)
& V(95.9\%) & V(95.8\%) & V(95.5\%)
& I(99.5\%) & I(99.5\%) & I(99.5\%)
& T(71.3\%) & T(62.8\%) & T(63.6\%) \\
\midrule

\multirow{3}{*}{WAVE}
& Hit@1 
& 0.0 & 66.3 & 0.0 
& 0.0 & 90.6 & 0.0 
& 0.0 & 0.0 & 0.0 
& 0.0 & 0.0 & 63.4 \\

& \cellcolor{gray!12}MRR
& \cellcolor{gray!12}2.6 & \cellcolor{gray!12}77.3 & \cellcolor{gray!12}2.7
& \cellcolor{gray!12}2.5 & \cellcolor{gray!12}94.4 & \cellcolor{gray!12}2.7
& \cellcolor{gray!12}2.5 & \cellcolor{gray!12}2.6 & \cellcolor{gray!12}2.7
& \cellcolor{gray!12}2.5 & \cellcolor{gray!12}2.6 & \cellcolor{gray!12}75.5 \\

& Top-10 DM
& V(99.9\%) & V(99.9\%) & V(99.9\%)
& V(99.9\%) & V(99.9\%) & V(99.9\%)
& V(99.9\%) & V(99.9\%) & V(99.9\%)
& V(99.9\%) & V(99.9\%) & V(99.9\%) \\
\midrule

\multirow{3}{*}{\makecell[c]{Qwen3-VL\\Embedding}}
& Hit@1 
& 0.0 & 0.0 & -- 
& 0.0 & 100.0 & -- 
& 0.0 & 2.5 & -- 
& -- & -- & -- \\

& \cellcolor{gray!12}MRR
& \cellcolor{gray!12}6.9 & \cellcolor{gray!12}2.9 & \cellcolor{gray!12}--
& \cellcolor{gray!12}6.6 & \cellcolor{gray!12}100.0 & \cellcolor{gray!12}--
& \cellcolor{gray!12}4.2 & \cellcolor{gray!12}15.9 & \cellcolor{gray!12}--
& \cellcolor{gray!12}-- & \cellcolor{gray!12}-- & \cellcolor{gray!12}-- \\

& Top-10 DM
& T(81.7\%) & T(84.0\%) & --
& V(97.8\%) & V(97.7\%) & --
& I(99.9\%) & I(99.9\%) & --
& -- & -- & -- \\
\bottomrule

\end{tabular}%
}
\caption{
Cross-modal retrieval performance across three models. DM denotes the dominant modality among the top-10 retrieved results.
}
\label{tab:cross_modal_all_models}
\end{table*}

\subsection{Explicit Modality Instructions Often Fail in Cross-Modal Retrieval}

\begin{tcolorbox}[
    colback=gray!5,
    colframe=black!60,
    boxrule=0.5pt,
    arc=1pt,
    boxsep=3pt,
    left=4pt,
    right=4pt,
    top=3pt,
    bottom=3pt
]
\noindent\textbf{Key Finding:}
Explicit modality instructions do not reliably lead to
correct target-modality retrieval, and cross-modal behavior
exhibits both asymmetry and modality bias.
\end{tcolorbox}
We first examine retrieval performance under explicit modality constraints. As shown in Table~\ref{tab:cross_modal_all_models}, most cross-modal directions remain challenging across all three models. In particular, Hit@1 is close to zero in many cases (e.g., T$\rightarrow$I, T$\rightarrow$A, V$\rightarrow$T), and even MRR remains low, indicating that the instructed target modality is often not retrieved. This suggests that modality-specific instructions alone are insufficient to consistently guide retrieval toward the desired modality.
A small number of directions achieve relatively high performance, such as I$\rightarrow$V and A$\rightarrow$T. However, these cases should be interpreted with caution. In OmniSET, videos are generated from images and audio from text, which introduces stronger intrinsic similarity for certain modality pairs. As a result, performance in these directions may partially reflect dataset construction effects rather than purely instruction-following capability. We provide a detailed discussion of this potential bias in Appendix~\ref{sec:synthetic_bias}.
Beyond overall performance, two consistent patterns emerge.

\textbf{(1) Cross-Modal Asymmetry.}
Cross-modal retrieval is highly directional. Strong performance in one direction does not imply comparable performance in the reverse. For example, I$\rightarrow$V achieves near-perfect scores (Hit@1 = 100.0 for Nemotron), while V$\rightarrow$I remains low (Hit@1 = 5.0). Similar asymmetry appears in other modality pairs (e.g., A$\rightarrow$T vs.\ T$\rightarrow$A), indicating that cross-modal relations are not bidirectionally aligned in the embedding space.
\textbf{(2) Modality bias.}
Retrieval results are dominated by modality proximity rather than target constraints. The Top-10 dominant modality (DM) is strongly correlated with the query modality across all models. For instance, Omni-Embed-Nemotron retrieves predominantly text for text queries (e.g., T2I: 84.9\%), while WAVE consistently retrieves video regardless of the target modality (99.9\% across all directions). Qwen3-VL shows a similar pattern, where retrieved modalities align more with the query modality than with the instructed target.

\subsection{Model Sensitivity to Modality-Constrained Instructions}

\begin{tcolorbox}[
    colback=gray!5,
    colframe=black!60,
    boxrule=0.5pt,
    arc=1pt,
    boxsep=3pt,
    left=4pt,
    right=4pt,
    top=3pt,
    bottom=3pt
]
\noindent\textbf{Key Finding}: Models vary substantially in their sensitivity to modality-constrained instructions, ranging from strong responsiveness to near invariance.
\end{tcolorbox}

\begin{figure*}[t]
    \centering
    \begin{subfigure}{0.32\textwidth}
        \centering
        \includegraphics[width=\linewidth]{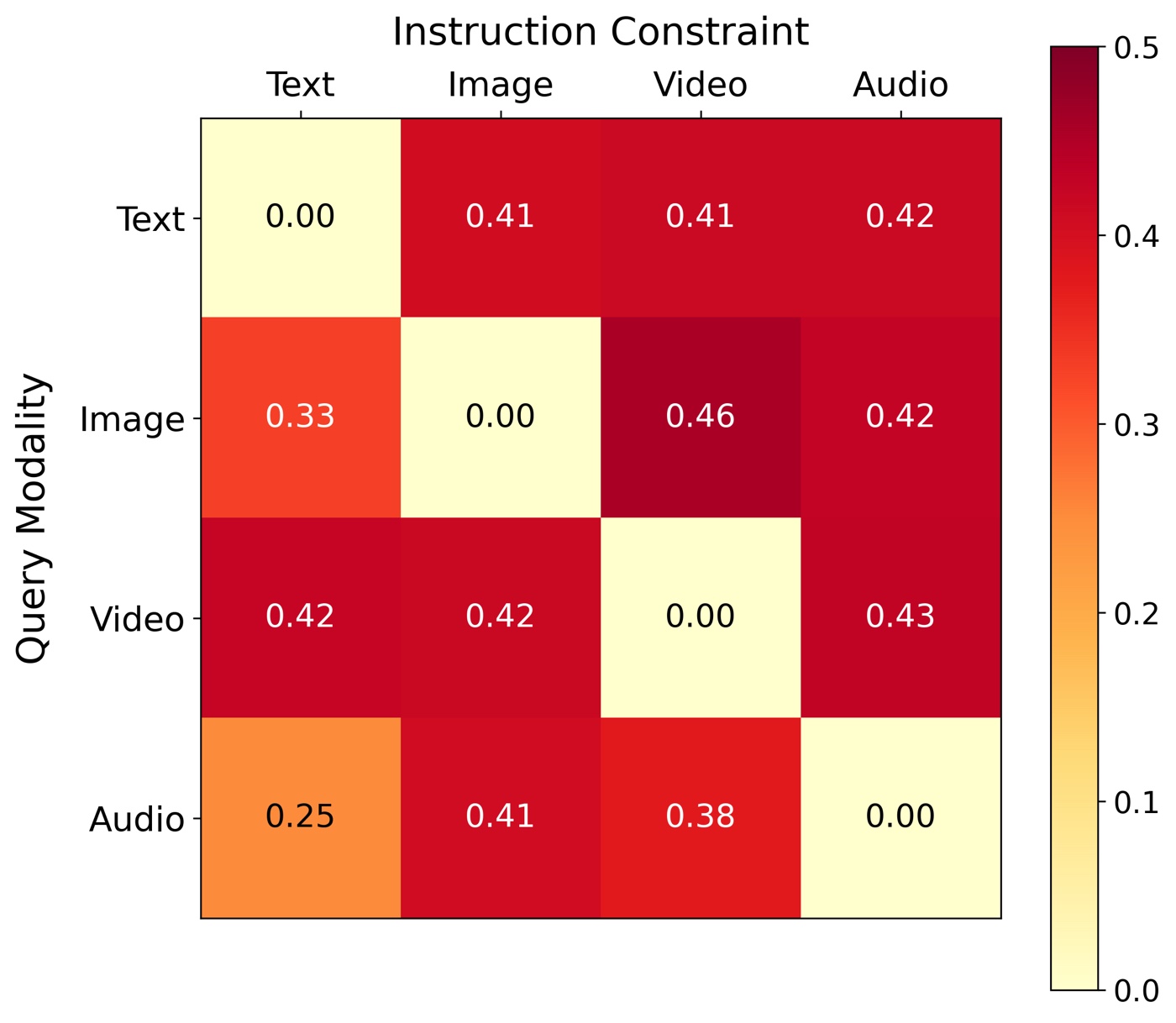}
        \caption{\texttt{Omni-Embed-Nemotron}.}
        \label{fig:instruction_shift_nemotron}
    \end{subfigure}
    \hfill
    \begin{subfigure}{0.32\textwidth}
        \centering
        \includegraphics[width=\linewidth]{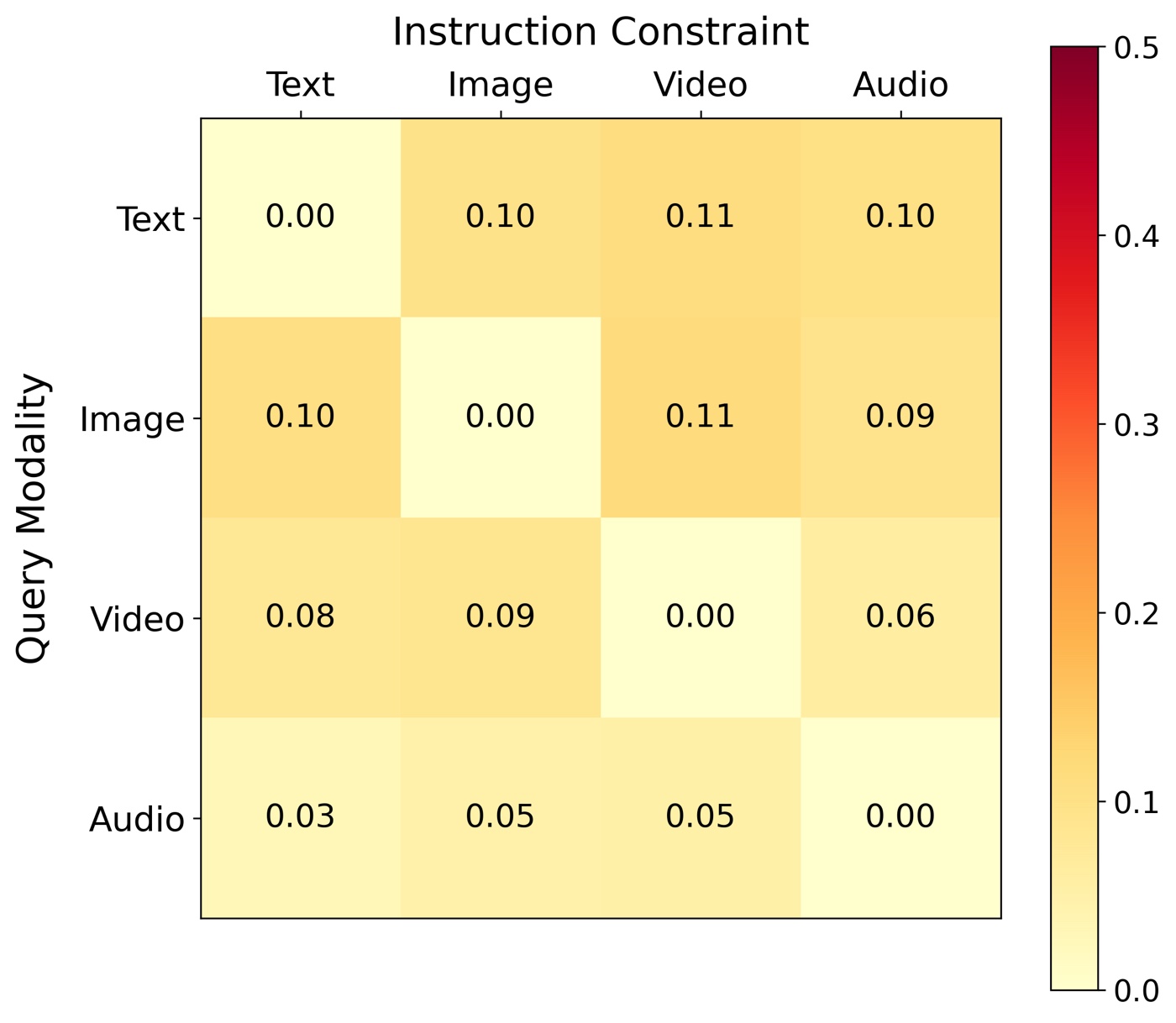}
        \caption{\texttt{WAVE}.}
        \label{fig:instruction_shift_wave}
    \end{subfigure}
    \hfill
    \begin{subfigure}{0.32\textwidth}
        \centering
        \includegraphics[width=\linewidth]{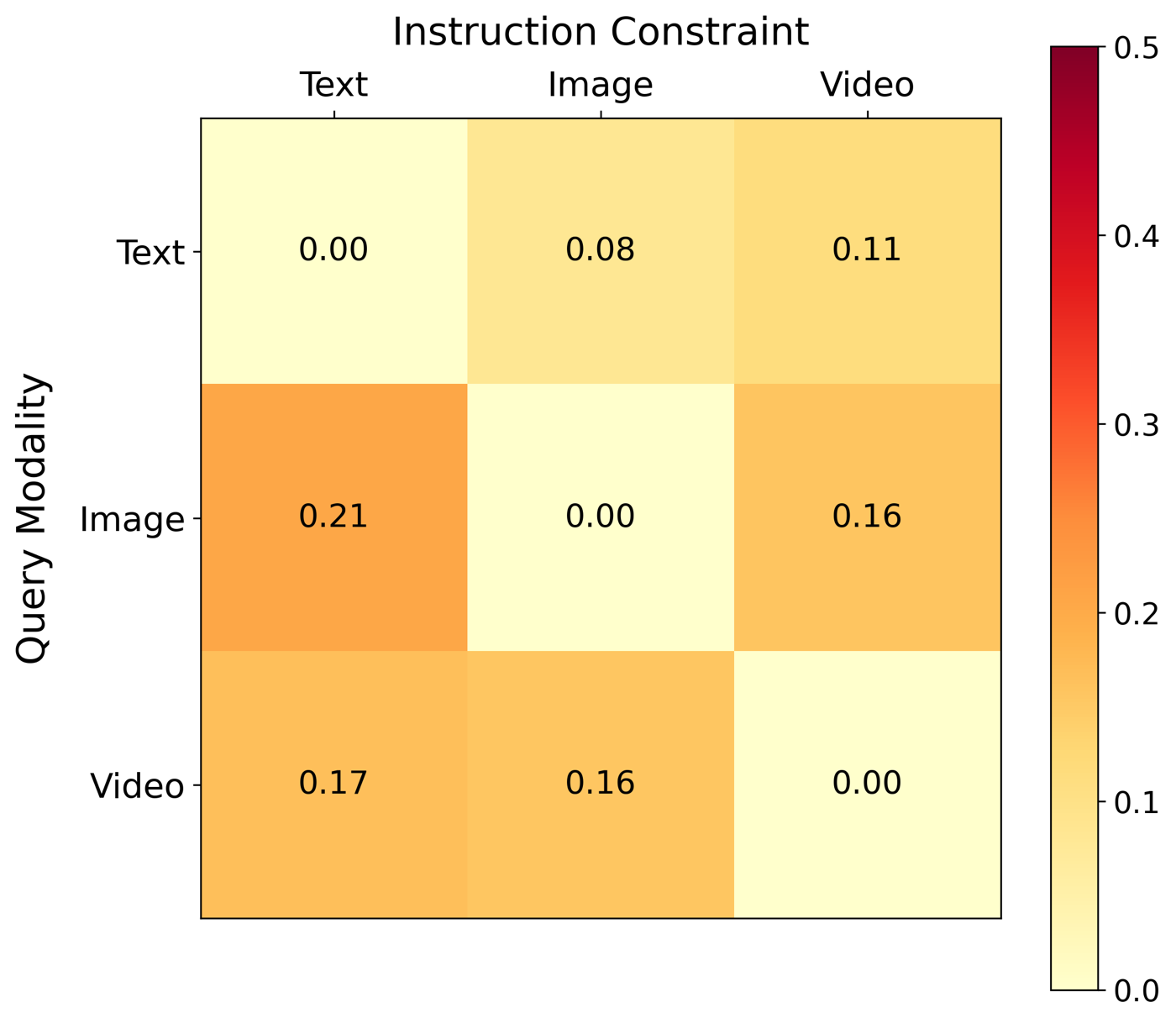}
        \caption{\texttt{Qwen3-VL-Embedding-8B}.}
        \label{fig:instruction_shift_qwen3}
    \end{subfigure}
\caption{
\textbf{Model sensitivity to modality-constrained instructions}, measured by the mean cosine distance between raw and instruction-augmented queries. Larger values indicate greater embedding shifts after instruction augmentation.
}
    \label{fig:instruction_shift_models}
\end{figure*}
To better understand the retrieval failures, we  examine how strongly each model responds to modality-constrained instructions at the representation level. We quantify this effect by measuring the cosine distance between the raw query embedding and its instruction-augmented counterpart, which captures the \emph{magnitude} of the instruction-induced shift.

Figure~\ref{fig:instruction_shift_models} shows clear differences across models. \texttt{Omni-Embed-Nemotron} is highly sensitive to modality-constrained instructions, with an average shift of about 0.4 cosine distance; for most cross-modal directions, the shift exceeds 0.3. In contrast, \texttt{WAVE} is much less responsive, with an average change of only 0.08, indicating that modality-constrained instructions have limited effect on its query representations. \texttt{Qwen3-VL-Embedding-8B} falls between these extremes: across text--image--video tasks, its average shift is about 0.15, suggesting moderate responsiveness.
These differences may be related to training objectives. Although Nemotron and WAVE both build on the Qwen2.5-Omni family, their responses to modality-constrained instructions differ substantially. While the instruction-related training details of Nemotron are not fully specified, its strong sensitivity and strong full-modality performance suggest that it likely benefited from more extensive instruction-related multimodal training. By contrast, WAVE focuses primarily on unified representation learning for visual and audio tasks, with training centered on multimodal retrieval and video QA rather than instruction-following. This difference in training emphasis may help explain its weaker response to modality-constrained instructions.
Overall, sensitivity to modality instructions is clearly not uniform across models. Some models barely react to such constraints, whereas others exhibit substantial embedding shifts. However, as we show next, stronger sensitivity does not necessarily imply better alignment with the target modality.

\subsection{Instruction-Induced Shifts Do Not Consistently Move Queries Toward the Target}

\begin{tcolorbox}[
    colback=gray!5,
    colframe=black!60,
    boxrule=0.5pt,
    arc=1pt,
    boxsep=3pt,
    left=4pt,
    right=4pt,
    top=3pt,
    bottom=3pt
]
\noindent\textbf{Key Finding:} Instruction-induced shifts do not consistently reduce the distance to the target modality, and often move queries farther away.
\end{tcolorbox}
\begin{figure*}[t]
    \centering
    \begin{subfigure}{0.35\textwidth}
        \centering
        \includegraphics[width=\linewidth]{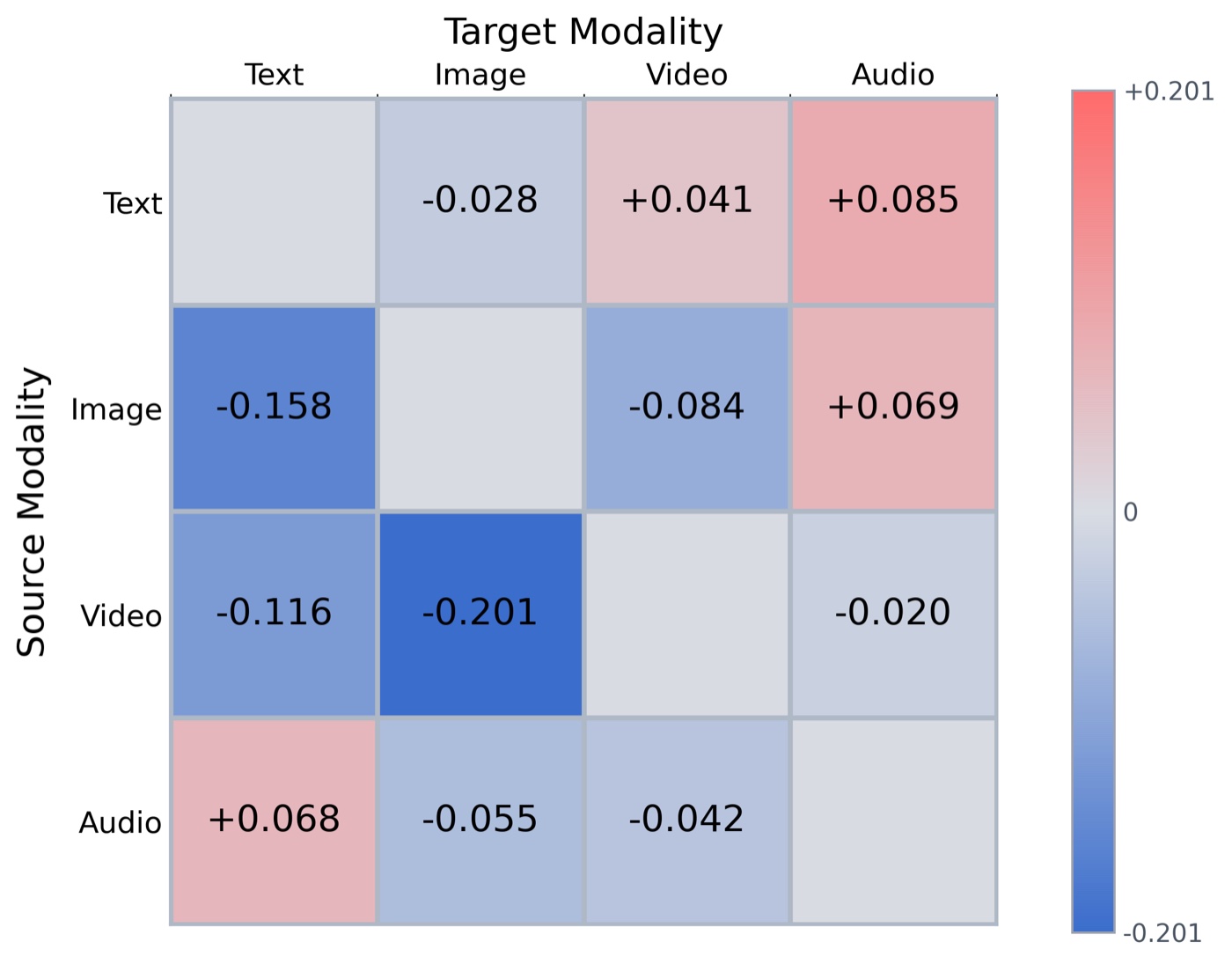}
        \caption{\texttt{Omni-Embed-Nemotron} (distance change heatmap).}
        \label{fig:target_shift_heatmap}
    \end{subfigure}
    \hfill
    \begin{subfigure}{0.31\textwidth}
        \centering
        \includegraphics[width=\linewidth]{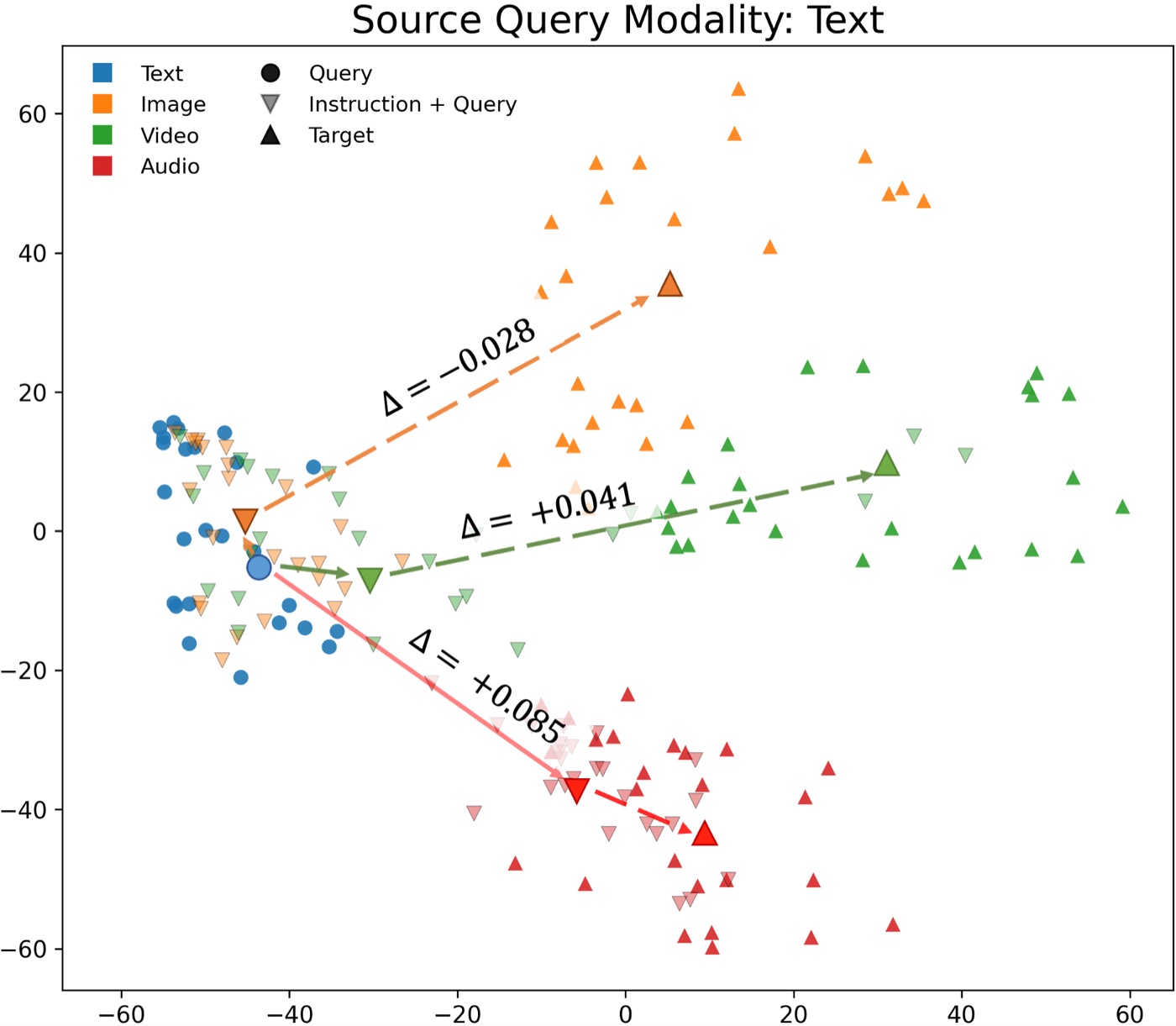}
        \caption{Text query (t-SNE visualization).}
        \label{fig:target_shift_text_via}
    \end{subfigure}
    \hfill
    \begin{subfigure}{0.31\textwidth}
        \centering
        \includegraphics[width=\linewidth]{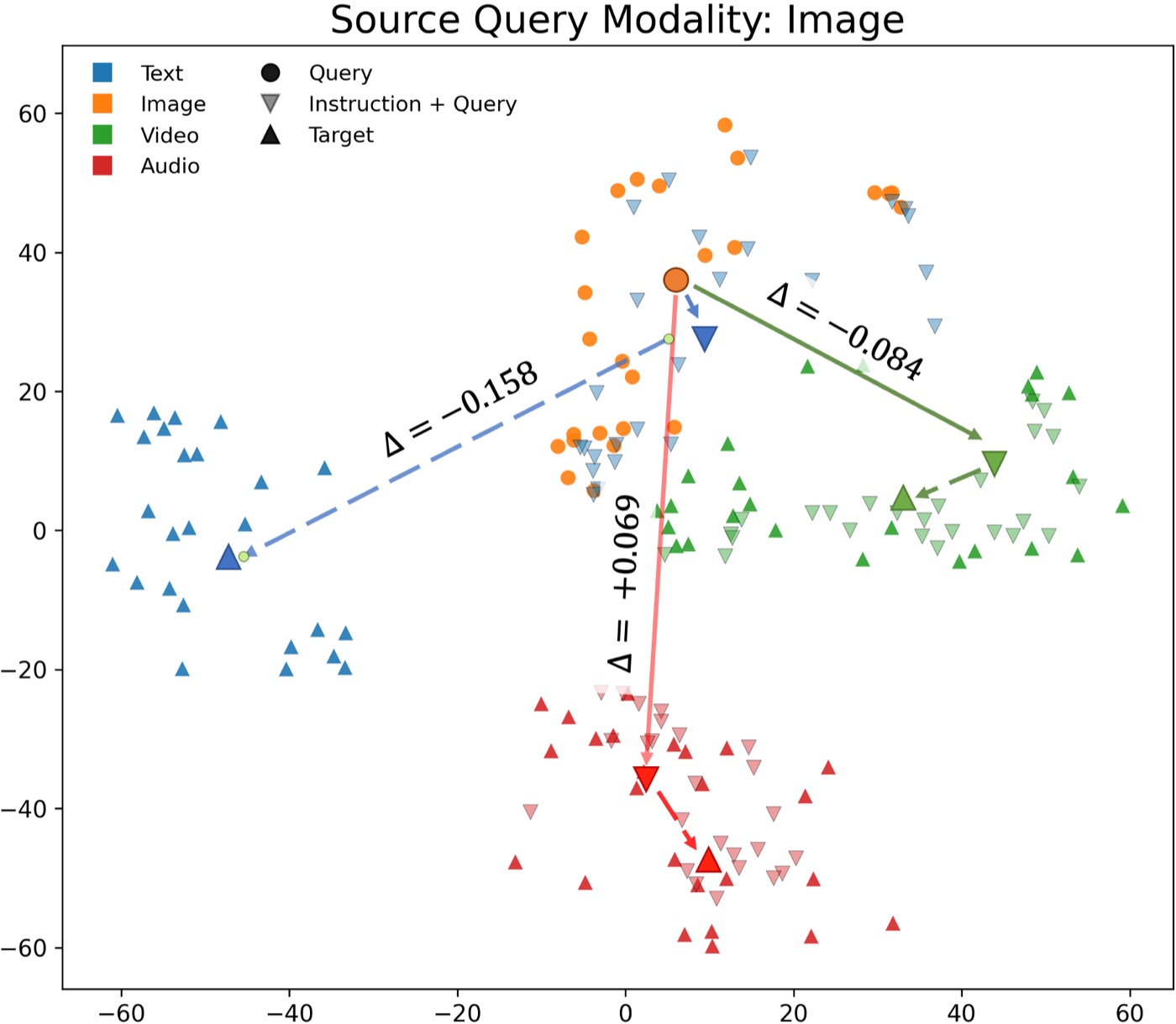}
        \caption{Image query (t-SNE visualization).}
        \label{fig:target_shift_image_vta}
    \end{subfigure}

\caption{
\textbf{Instruction-induced changes in query--target distance and embedding shifts.}
(a) Change in cosine distance to the target modality after instruction augmentation, measured relative to the raw query.
(b,c) t-SNE visualizations showing how instruction-augmented queries move in the embedding space for text and image queries.
In (b,c), raw queries are shown as circles (\scalebox{1.8}{$\bullet$}), instruction-augmented queries as downward triangles ($\blacktriangledown$), and target instances as upward triangles ($\blacktriangle$).
}
\label{fig:instruction_target_shift_analysis}
\end{figure*}
We next analyze the \emph{direction} of instruction-induced shifts, i.e., whether instruction augmentation moves queries closer to the intended target modality. We compare the distance from the raw query to the target with that from the instruction-augmented query to the same target.
We focus on \texttt{Omni-Embed-Nemotron}, which exhibits the strongest sensitivity to modality-constrained instructions; visualizations for other models are provided in Appendix~\ref{sec:visualiztion}.
As shown in Figure~\ref{fig:target_shift_heatmap}, instruction-induced shifts do not consistently improve alignment. For Nemotron, only a few directions (e.g., T$\rightarrow$V, T$\rightarrow$A, A$\rightarrow$T, I$\rightarrow$A) show slight improvements, all below 0.09, while most directions instead increase the distance to the target modality. 
Moreover, this behavior is \emph{asymmetric} across directions. For instance, T$\rightarrow$V exhibits a small improvement (+0.041), whereas the reverse direction V$\rightarrow$T shows a much larger degradation ($-0.158$). This asymmetry suggests that instruction-induced shifts do not establish a consistent bidirectional alignment between modality pairs, but instead interact unevenly with the underlying embedding geometry.
The t-SNE visualizations in Figures~\ref{fig:target_shift_text_via} and~\ref{fig:target_shift_image_vta} provide qualitative evidence. Although instruction augmentation induces noticeable movement in the embedding space, these shifts are not consistently oriented toward the target modality and can deviate toward other modality clusters. For example, image queries augmented with a text-target instruction are often observed to move closer to the video cluster rather than the text cluster.
A similar trend is observed for \texttt{Qwen3-VL-Embedding-8B} (see Appendix~\ref{sec:visualiztion}), where instruction-induced shifts yield only limited improvements in a small number of directions, while most changes do not reduce the query--target distance.
Overall, instruction-induced shifts do not reliably translate into target-oriented movement. Even when models are sensitive to modality instructions, the induced changes are not consistently aligned with the target modality, limiting their effectiveness for cross-modal retrieval.

\section{Conclusion}
In this work, we advance the evaluation of multimodal embeddings toward the \textit{full-modality} setting. We introduce MMEB-V3, a comprehensive benchmark that systematically evaluates embeddings across text, image, video, audio, and agent-centric scenarios. To enable controlled analysis, we further construct Omni-modality Semantic Equivalence Tuples (OmniSET), which isolate modality effects under consistent semantics and enable rigorous evaluation of cross-modal behavior under instruction constraints.
Based on this framework, we conduct a systematic analysis of modality-constrained retrieval and identify several key limitations. We find that models often fail to retrieve the intended target modality, cross-modal retrieval is highly asymmetric and exhibits strong directional biases, and instruction-conditioned signals do not reliably guide retrieval toward the desired modality.
These findings suggest that current multimodal embeddings remain limited in supporting reliable modality-aware retrieval under instruction constraints. We hope MMEB-V3 provides a useful benchmark for diagnosing these limitations and for guiding future research toward more controllable full-modality embedding models, particularly in emerging agent-centric applications.

\section*{Ethics Statement}
We develop and evaluate general-purpose multimodal retrieval and embedding systems using publicly available datasets. No private or personally identifiable data is used. While our benchmarks may inherit biases from underlying corpora, we do not explicitly address fairness, and such biases may affect downstream applications. The proposed methods are intended for beneficial uses such as information access, but may have dual-use risks if misapplied. We encourage responsible use, careful dataset curation, and transparency through the release of code and evaluation protocols.

\bibliographystyle{plainnat}
\bibliography{references}

\appendix
\section{Appendix}
\subsection{Details of Baseline Models}

\textbf{Omni-Embed-Nemotron}~\citep{xu2025omniembednemotronunifiedmultimodalretrieval} is a unified multimodal embedding model designed for retrieval over text, images, audio, and video. Built on the Thinker component of Qwen2.5-Omni-3B, it encodes different modalities into a shared embedding space and is trained with a bi-encoder contrastive objective for multimodal retrieval and RAG scenarios.

\textbf{WAVE}~\citep{tang2025wavelearningunified} is a unified audio-visual embedding model for cross-modal retrieval and multimodal understanding. It learns shared audio-visual representations through joint multimodal training and hierarchical feature fusion. Combined with unified embedding learning, it achieves strong performance on audio-visual retrieval and multimodal question answering tasks.

\textbf{Qwen3-VL-Embedding}~\citep{qwen3vlembedding} is a multimodal embedding model built on the Qwen3-VL foundation model for cross-modal retrieval and understanding. It supports text, images, screenshots, videos, and mixed-modal inputs within a unified framework, learning unified representations across modalities and languages. Using dense embedding representations, it enables high-quality multimodal retrieval and clustering.

\textbf{VLM2Vec-V2.0}~\citep{meng2025vlm2vecv2advancingmultimodalembedding} is a multimodal embedding model that extends unified representation learning to diverse visual modalities. It supports images, videos, and visual documents within a shared embedding space, and employs unified training across modalities to learn consistent representations. Using contrastive embedding learning, it enables effective retrieval in diverse multimodal settings.

\textbf{VLM2Vec}~\citep{jiang2024vlm2vec} converts an instruction-tuned vision-language model into a unified multimodal embedding model for retrieval. It reformulates multimodal inputs into embedding-oriented representations under a unified framework, enabling consistent encoding across modalities. Using contrastive embedding learning, it achieves strong performance across diverse multimodal retrieval tasks.

\textbf{GME}~\citep{zhang2025gme} is a unified multimodal embedding model based on Qwen2-VL. It supports three input types—text, image, and image-text pairs—and maps both single-modal and combined-modal inputs into universal vector representations, enabling versatile Any2Any retrieval scenarios such as text-to-image, image-to-image, and multimodal search.

\subsection{Details of Benchmark Construction}

\textbf{MMEB-V3} represents a significant advancement over its predecessor, MMEB-V2, by establishing \textit{a comprehensive unified embedding evaluation framework} that encompasses text, image, video, and audio modalities. While previous iterations focused predominantly on static images and eventually expanded to videos and visual documents, MMEB-V3 fills critical gaps in the embedding landscape by integrating comprehensive audio support, complex text reasoning, and specialized agentic capabilities. In addition, we further enrich the image domain by introducing a multi-condition, multimodal retrieval dataset, \textbf{MCMR}~\citep{MCMR}, which evaluates fine-grained cross-modal matching under multiple constraints. By consolidating 190 heterogeneous tasks into a standardized ranking-based evaluation, MMEB-V3 provides a rigorous testbed for developing general-purpose, omni-modality embeddings capable of instruction-following and cross-modal semantic alignment.

\paragraph{Audio Tasks.}
To address the historical scarcity of audio-focused evaluation in unified models, MMEB-V3 introduces a suite of audio tasks covering classification, cross-modal retrieval, and temporal grounding.  For several large-scale audio datasets, including NSynth, SpeechCommands, SoundDescs, and SpeechCOCO, we subsample up to 1,000 queries and limit the candidate pool to at most 10,000 instances. This strategy ensures computational tractability while maintaining a representative evaluation of model performance, consistent with common practices in large-scale retrieval benchmarks:

\begin{itemize}[leftmargin=*]

\item \textbf{Audio Classification.} 
Datasets include ESC-50~\citep{esc50}, UrbanSound8K~\citep{urbansound}, NSynth~\citep{nsynth2017}, Speech Commands~\citep{speechcommands}, and CREMA-D~\citep{cremad}, evaluating the ability to recognize discrete acoustic events and sound categories.

\item \textbf{Cross-modal Audio Retrieval.} 
This task evaluates alignment between audio and other modalities, including text-to-audio retrieval with Clotho~\citep{2020clotho} and SoundDescs~\citep{sounddescs}, audio-video alignment with AVE~\citep{ave}, and audio-image retrieval with SpeechCOCO~\citep{speechcoco}.

\item \textbf{Audio Temporal Grounding.} 
Using the TUT Sound Events 2017 dataset~\citep{tutsound2017}, models must localize specific acoustic events within continuous audio streams.

\end{itemize}

\paragraph{Text Tasks.}
Recognizing that standard text retrieval tasks often fail to challenge modern information retrieval systems~\citep{lu2026rethinking}, MMEB-V3 introduces more demanding scenarios involving instruction following, reasoning, and long-context understanding:

\begin{itemize}[leftmargin=*]

\item \textbf{Instruction Following Retrieval.} 
FollowIR~\citep{weller-etal-2025-followir} and InfoSearch~\citep{zhou2025beyond} evaluate whether models can retrieve documents satisfying complex instructions and constraints.

\item \textbf{Reasoning Retrieval.} 
BRIGHT~\citep{su2025bright} and R2MD~\citep{li2025r2medbenchmarkreasoningdrivenmedical} require logical inference and multi-hop reasoning beyond simple keyword matching.

\item \textbf{Long-context Retrieval.} 
LongEmb~\citep{zhu-etal-2024-longembed} evaluates retrieval from long documents and extended contexts.

\item \textbf{Multi-condition Retrieval.} 
MultiConIR~\citep{lu-etal-2025-multiconir} measures the ability to satisfy multiple retrieval constraints simultaneously.

\item \textbf{General Text Retrieval.} 
nanoBEIR~\citep{thakur2021beirheterogenousbenchmarkzeroshot}, a curated subset of BEIR, provides a compact yet diverse benchmark for semantic retrieval.

\end{itemize}

\paragraph{Agent Tasks.}
MMEB-V3 further evaluates agentic capabilities such as tool selection, GUI interaction and agent memory retrieval:

\begin{itemize}[leftmargin=*]

\item \textbf{Tool Retrieval.} 
Tool-REX~\citep{lu2025toolsunderdocumentedsimpledocument} contains over 43,000 tools (e.g., APIs and code functions) with structured metadata, evaluating the ability to retrieve appropriate tools for user intents.

\item \textbf{GUI Agent Trajectory Retrieval.} 
GAE-Bench~\citep{gui} evaluates retrieval over multimodal GUI interaction trajectories represented by screenshots and structured actions.

\item \textbf{Agent Memory Retrieval.} 
LMEB~\citep{zhao2026lmeblonghorizonmemoryembedding} is a large-scale benchmark suite for long-horizon memory modeling. We select four representative tasks covering diverse memory types: \emph{episodic memory} (KnowMeBench~\citep{wu2026knowmebenchbenchmarkingpersonunderstanding}), \emph{dialogue memory} (REALTALK~\citep{lee2025realtalk21dayrealworlddataset}), \emph{semantic memory} (PeerQA~\citep{baumgärtner2025peerqascientificquestionanswering}), and \emph{procedural memory} (DeepPlanning~\citep{zhang2026deepplanningbenchmarkinglonghorizonagentic}). These tasks are used to evaluate whether embedding models can retrieve and utilize relevant memory under diverse semantic structures and temporal dependencies.

\end{itemize}
\paragraph{Omni-modality Semantic Equivalence Tuples}
Cross-modal retrieval is an information retrieval setting where a query in one modality is used to retrieve semantically aligned instances from a different modality. Despite recent progress in omni-embedding models that learn shared representation spaces across modalities, existing evaluations remain limited in measuring fine-grained, modality-aware alignment. To address this, MMEB-V3 introduces the concept of \textbf{Omni-modality Semantic Equivalence Tuples}, which is defined as a set of aligned instances $\{x^T,x^I,x^V,x^A\}$ with each instance encoding the same semantic meaning in one of the four different modalities. Under this setting, cross-modal retrieval is defined as the task of retrieving the correct element from a shared (flattened) pool of OmniSET that (1) is semantically equivalent to the input instance and (2) belongs to the specified target modality in the instruction.

\begin{itemize}[leftmargin=*]
\item \textbf{OmniSET} contains 100 queries with human-verified hard negatives to increase discriminative difficulty. Each query is further expanded into 12 directional tasks, corresponding to the 12 ordered input-target modality pairs $(m_{in} \rightarrow m_{target})$ where $m_{in} \neq m_{target}$ and $m \in \{T,I,V,A\}$. 
\end{itemize}

\subsubsection{Construction of OmniSET}
\label{sec: cmet construction}

We introduce \textbf{OmniSET (Omni-modality Semantic Equivalence Tuples)}, a dataset designed to construct semantically equivalent instances across different modalities. The goal is to create tuples that share the same underlying semantics while differing in modality, enabling controlled evaluation of cross-modal alignment and transfer.
Built upon the widely used MSCOCO dataset~\citep{MSCOCO}, which provides diverse real-world images with human-annotated captions, we extend the original bimodal (image--text) setting into a unified four-modality framework. Specifically, for each selected instance, we construct an \emph{Omni-modality Semantic Equivalence Tuples (OmniSET)} consisting of aligned representations in text, image, video, and audio modalities.

\begin{figure}[ht]
    \centering
    \includegraphics[width=0.99\linewidth]{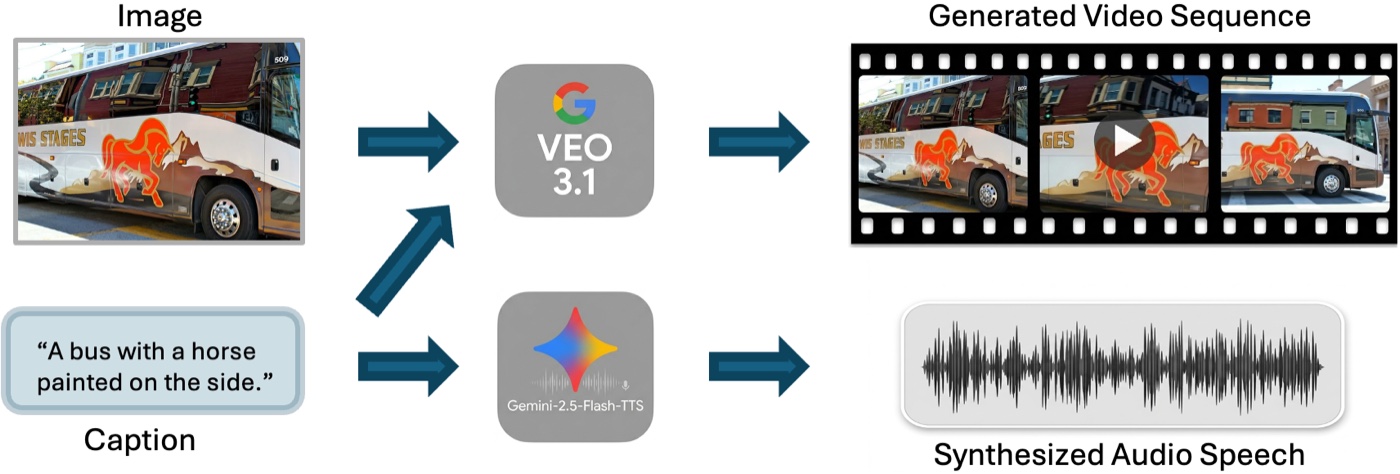}
    \caption{Construction pipeline of OmniSET.}
    \label{fig:cmet_pipeline}
\end{figure}

Our construction consists of two main components. First, we curate a set of approximately 200 high-quality query samples, each paired with human-verified hard negatives to ensure fine-grained semantic discrimination. Second, we augment the original modalities by synthesizing additional ones: motion videos are generated from images using Google Veo-3.1, and speech audio is generated from captions using Gemini-2.5-Flash-TTS. This results in semantically aligned multi-modal tuples with consistent content across modalities.
The overall construction pipeline is illustrated in Figure~\ref{fig:cmet_pipeline}, and the prompt templates used for video and audio generation are provided in Figure~\ref{fig:generation-prompts}.

The hard negative construction process is performed as follows:
\begin{enumerate}[leftmargin=*, itemsep=2pt]
    \item We filter out images with restrictive licenses (e.g., Attribution-NoDerivs and Attribution-NonCommercial-NoDerivs) that prohibit derivative content.
    \item For each remaining image, we extract object annotations and compute the number of unique object categories.
    \item Given a reference image, we rank all other images by the number of shared object categories and select the top 30 as candidate hard negatives.
    \item We sample approximately 100 images as query instances.
    \item We manually inspect the candidates to remove near-duplicates or overly similar samples, resulting in 15--20 hard negatives per query.
    \item Finally, for each positive instance and its associated negatives, we generate videos and audio to form complete OmniSET across all four modalities.
\end{enumerate}


\begin{figure}[ht]
    \centering
    \begin{mdframed}[
        linewidth=0.8pt,
        linecolor=gray!60,
        backgroundcolor=gray!10,
        roundcorner=4pt,
        innertopmargin=6pt,
        innerbottommargin=6pt,
        innerleftmargin=8pt,
        innerrightmargin=8pt,
        tikzsetting={dashed},
    ]
\small
\textbf{Video Generation Prompt}
\vspace{2pt}

\begin{lstlisting}
Generate a video based on the following image and description:
[Reference Image] + [caption]
\end{lstlisting}

\vspace{4pt}
\textbf{Audio Generation Prompt}
\vspace{2pt}

\begin{lstlisting}
Read aloud the following sentence in a natural and expressive way,
in a warm and friendly tone:
[caption]
\end{lstlisting}
    \end{mdframed}
    \caption{Prompt templates for video and audio generation.}
    \label{fig:generation-prompts}
\end{figure}

We evaluate cross-modal retrieval by formulating queries and targets in different modalities. Each query is provided in one modality, and the model is required to retrieve a semantically matching instance from a specified target modality.

Given four modalities (text, image, video, and audio), each modality can query the other three, resulting in a total of 12 directed retrieval tasks (e.g., image $\rightarrow$ text, image $\rightarrow$ video, image $\rightarrow$ audio).

\paragraph{Shared candidate pool.}
For each query, all 12 retrieval directions share the same candidate pool, which contains the corresponding hard negatives across all four modalities. Therefore, regardless of the query direction, the model must select the correct target from a unified mixed-modality pool. This design ensures that performance differences across directions are not due to variations in candidate sets.

\paragraph{Avoiding trivial matching.}
A limitation of our dataset construction is that generated videos are highly consistent with their source images, which may simplify certain directions such as I$\rightarrow$V and V$\rightarrow$I. To mitigate this, the input instance is explicitly included in the candidate pool as a distractor, while same-modality retrieval (e.g., I$\rightarrow$I and V$\rightarrow$V) is excluded. This prevents trivial matching based on instance identity.

\subsubsection{Impact of Synthetic Data and Potential Modality Bias}
\label{sec:synthetic_bias}

A key design choice in OmniSET is the use of synthetic data to complete modality coverage: videos are generated from images, and audio is generated from text captions. While this enables controlled construction of semantically aligned tuples across modalities, it may introduce systematic biases that affect certain cross-modal directions.

In particular, the generation process creates asymmetric dependencies between modalities. Video samples are directly derived from images, and audio samples are derived from text. As a result, modality pairs such as image--video and text--audio may exhibit higher intrinsic similarity than other cross-modal pairs. This effect can partially explain the unusually strong performance observed in directions such as I$\rightarrow$V and A$\rightarrow$T, where retrieval may benefit from shared generation artifacts rather than purely learned cross-modal alignment.

More broadly, this construction may introduce a form of \emph{modality preference}, where certain modality pairs are more tightly coupled due to data generation rather than model capability. This could potentially amplify observed behaviors such as modality bias or directional asymmetry. For example, when a query modality is closely aligned with a generated modality, retrieval may favor that modality even when it is not explicitly specified as the target.

However, several aspects of our design mitigate the extent to which these biases affect our conclusions. First, all retrieval directions share a unified mixed-modality candidate pool, ensuring that performance differences are not caused by variations in candidate sets. Second, trivial matching is explicitly avoided by including the source instance as a distractor and excluding same-modality retrieval. Third, the key phenomena observed in our analysis—such as the failure of instruction-constrained retrieval and the misalignment of instruction-induced shifts—are consistent across multiple models and modality directions, including those not directly affected by synthetic generation.

We therefore view these biases as an inherent trade-off of controlled multi-modal construction: while synthetic data may strengthen certain modality pairs, it does not fully account for the broader patterns observed in our experiments. Nevertheless, future work could further investigate this issue by incorporating human-annotated multi-modal data or by disentangling synthetic and real modality pairs through targeted ablation studies.


\begin{figure*}[th]
    \centering
    \includegraphics[width=0.9\linewidth]{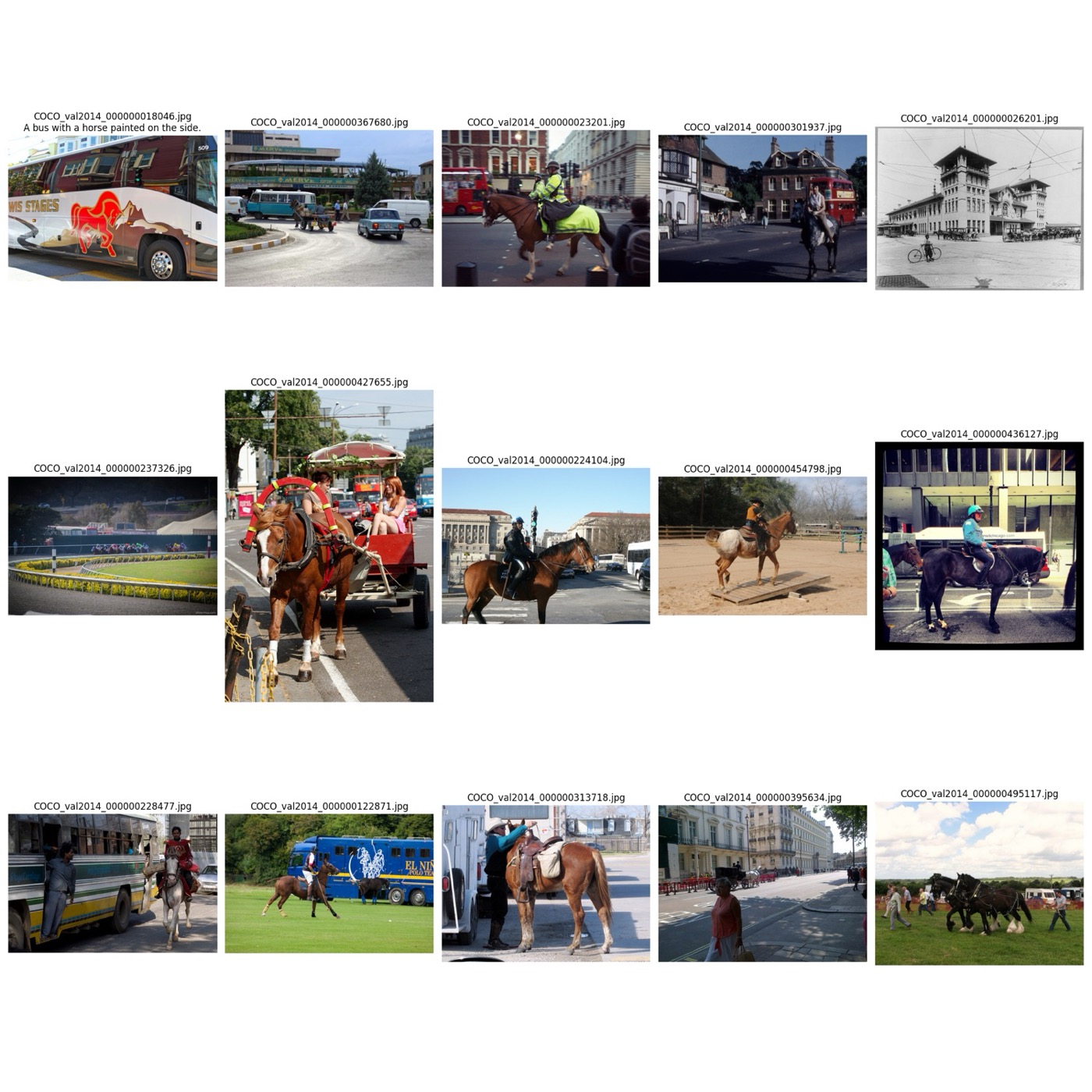}
    \caption{Example Hard Negative Set 18046 from OmniSET.  \textbf{Query 
    Image Caption}: A \textbf{bus} with a \textbf{horse printed} on the side. \textit{*The first image is the query input image}}
    \label{fig:sample_18046_mscoco-cmet}
\end{figure*}
\begin{figure*}[th]
    \centering
    \includegraphics[width=0.9\linewidth]{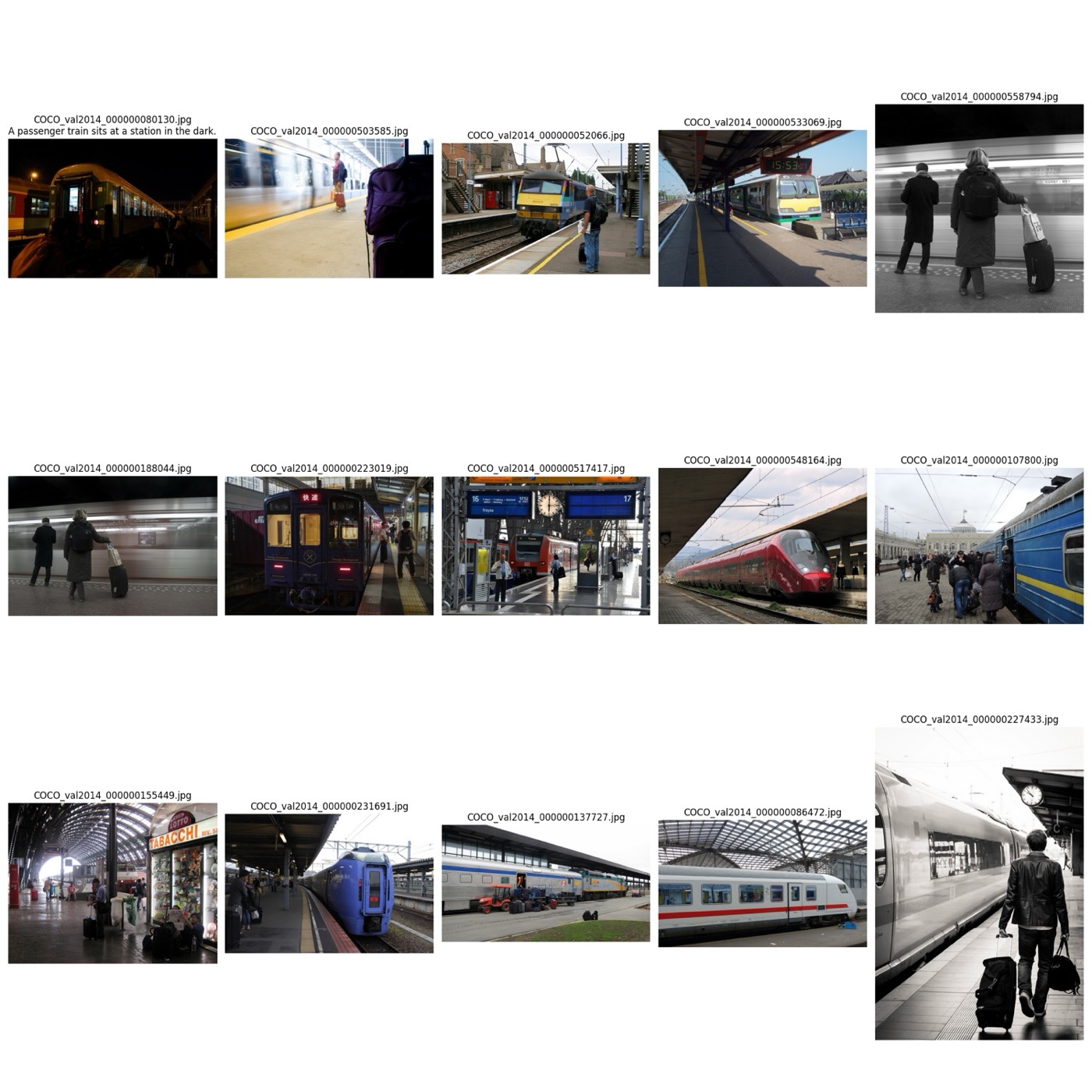}
    \caption{Example Hard Negative Set 80130 from OmniSET. \textbf{Query 
    Image Caption}: A passenger train \textbf{sits} at a station \textbf{in the dark}. \textit{*The first image is the query input image}}
    \label{fig:sample_80130_mscoco-cmet}
\end{figure*}
\begin{figure*}[th]
    \centering
    \includegraphics[width=0.9\linewidth]{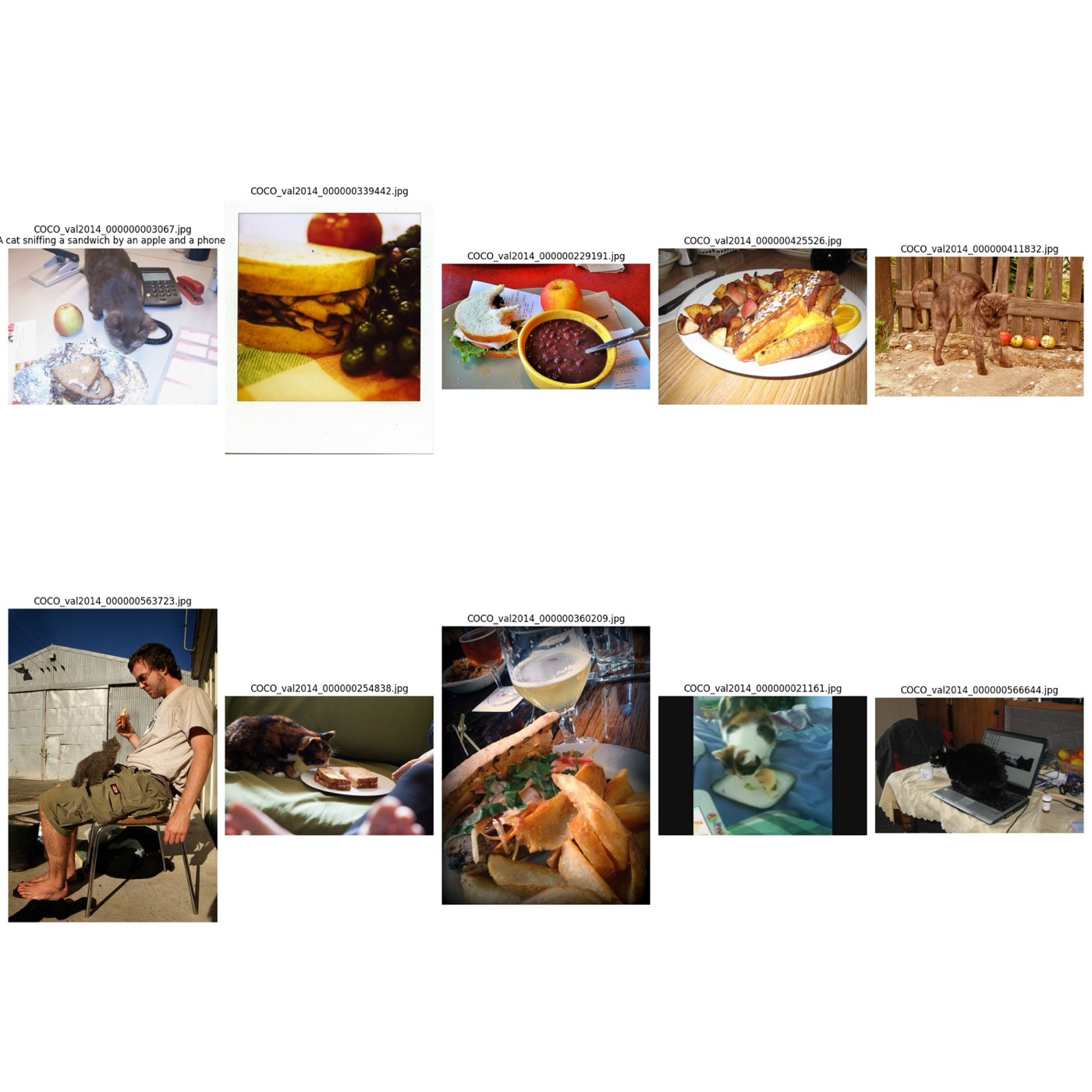}
    \caption{Example Hard Negative Set 3067 from OmniSET. \textbf{Query 
    Image Caption}: A \textbf{cat} sniffing a \textbf{sandwich} by an \textbf{apple} and a \textbf{phone}. \textit{*The first image is the query input image}}
    \label{fig:sample_3067_mscoco-cmet}
\end{figure*}

\subsection{Detailed Scores}
\label{sec: detailed scores}

We provide complete per-dataset evaluation results for each task .

\textbf{Image tasks.} Results are reported in Table~\ref{tab:image_appendix_results}.

\textbf{Video tasks.} Results are reported in Table~\ref{tab:video_appendix_results}.

\textbf{VisDoc tasks.} Results are reported in Table~\ref{tab:visdoc_appendix_results}.

\textbf{Audio tasks.} Results are reported in Table~\ref{tab:audio_appendix_results}.

\textbf{Text tasks.} Results are reported in Table~\ref{tab:text_appendix_results}.

\textbf{Agent tasks.} Results are reported in Table~\ref{tab:agent_appendix_results}.

\begin{table*}[t]
\centering
\scriptsize
\setlength{\tabcolsep}{3pt}
\renewcommand{\arraystretch}{0.95}
\resizebox{\textwidth}{!}{%
\begin{tabular}{l|cccccccccc}
\toprule
\textbf{Dataset}
& \textbf{Qwen3-VL}
& \textbf{Qwen3-VL}
& \textbf{VLM2Vec}
& \textbf{VLM2Vec}
& \textbf{GME}
& \textbf{WAVE}
& \textbf{Omni-Embed}
& \textbf{LCO-Embedding}
& \textbf{E5-Omni}
& \textbf{E5-Omni} \\
& \textbf{-Embedding(2B)}
& \textbf{-Embedding(8B)}
& \textbf{-Qwen2VL (7B)}
& \textbf{-V2.0 (2B)}
& \textbf{(7B)}
& \textbf{(7B)}
& \textbf{-Nemotron (3B)}
& \textbf{-Omni (7B)}
& \textbf{(3B)}
& \textbf{(7B)} \\
\midrule

\rowcolor{blue!10}
\textbf{Avg - Image (37 tasks, Hit@1)} &69.5 &72.1 &63.6 &63.3 &55.2 &41.5 &43.4 &49.6 &63.5 & 70.5  \\

\midrule

\rowcolor{orange!15}
\textbf{I-CLS (10)} & 63.6 & 65.2 & 62.8 & 62.9 & 57.6 & 50.1 & 48.3 & 57.3 & 60.7 & 68.2 \\
\rowcolor{orange!15}
\textbf{I-QA (10)} & 71.5 & 76.8 & 56.5 & 56.4 & 34.6 & 26.3 & 19.9 & 39.5 & 62.2 & 70.3 \\
\rowcolor{orange!15}
\textbf{I-RET (13)} & 68.4 & 69.0 & 69.4 & 69.6 & 71.2 & 42.1 & 56.1 & 48.8 & 65.0 & 69.7 \\
\rowcolor{orange!15}
\textbf{I-VG (4)} & 82.9 & 87.7 & 81.9 & 77.1 & 59.5 & 56.1 & 48.9 & 58.1 & 68.6 & 79.6 \\

\midrule

ImageNet-1K & 71.2 & 75.3 & 80.2 & 80.8 & 64.7 & 43.8 & 61.6 & 69.9 & 72.0 & 79.9 \\
N24News & 65.2 & 64.4 & 79.5 & 73.0 & 50.3 & 43.7 & 36.7 & 37.4 & 73.2 & 77.9 \\
HatefulMemes & 59.7 & 68.4 & 70.3 & 55.8 & 53.9 & 51.1 & 48.8 & 46.8 & 56.6 & 69.9 \\
VOC2007 & 84.1 & 84.8 & 80.6 & 84.9 & 80.1 & 79.9 & 53.2 & 74.5 & 75.2 & 84.5 \\
SUN397 & 67.9 & 64.7 & 77.3 & 70.9 & 69.4 & 64.1 & 55.1 & 70.2 & 71.0 & 76.0 \\
Place365 & 37.8 & 37.6 & 37.2 & 36.1 & 39.1 & 34.0 & 31.9 & 39.0 & 41.0 & 44.5 \\
ImageNet-A & 59.9 & 61.6 & 57.6 & 47.6 & 40.6 & 25.2 & 42.0 & 50.7 & 42.3 & 60.5 \\
ImageNet-R & 91.1 & 92.0 & 74.2 & 89.3 & 83.9 & 69.1 & 82.8 & 91.0 & 85.2 & 86.6 \\
ObjectNet & 78.7 & 78.2 & 40.5 & 65.1 & 69.0 & 69.1 & 46.6 & 66.2 & 64.7 & 69.8 \\
Country211 & 20.7 & 24.5 & 30.2 & 25.8 & 24.6 & 21.0 & 24.4 & 27.2 & 25.9 & 32.3 \\
OK-VQA & 65.9 & 74.7 & 56.8 & 51.7 & 33.1 & 32.7 & 17.7 & 46.7 & 62.7 & 70.6 \\
A-OKVQA & 58.8 & 66.6 & 47.5 & 44.0 & 20.8 & 24.8 & 12.5 & 39.5 & 52.3 & 59.7 \\
DocVQA & 93.9 & 95.2 & 88.8 & 90.1 & 41.1 & 19.7 & 17.1 & 80.7 & 86.6 & 93.6 \\
InfographicsVQA & 69.5 & 81.5 & 59.0 & 59.1 & 20.5 & 16.9 & 8.5 & 7.4 & 55.3 & 72.7 \\
ChartQA & 61.4 & 70.4 & 56.6 & 48.1 & 17.7 & 13.3 & 13.6 &12.5 &48.2 &65.1 \\
Visual7W & 62.8 & 65.9 & 52.7 & 52.8 & 22.2 & 19.2 & 7.3 & 9.3 & 55.7 & 63.6 \\
ScienceQA & 71.4 & 75.6 & 38.2 & 38.1 & 28.2 & 21.6 & 24.8 & 22.6 & 47.9 & 56.4 \\
VizWiz & 57.2 & 59.5 & 39.6 & 43.3 & 38.9 & 33.1 & 34.2 & 31.1 & 51.8 & 56.1 \\
GQA & 85.9 & 88.6 & 54.4 & 65.4 & 76.8 & 50.6 & 28.0 & 69.8 & 78.5 & 78.9 \\
TextVQA & 88.1 & 90.2 & 71.6 & 71.6 & 46.4 & 30.6 & 35.3 & 74.9 & 82.8 & 86.3 \\
VisDial & 74.8 & 66.8 & 81.6 & 82.7 & 60.9 & 21.8 & 51.3 & 51.0 & 77.8 & 80.2 \\
CIRR & 55.5 & 56.7 & 51.4 & 57.3 & 54.9 & 31.9 & 13.2 & 8.1 & 36.8 & 49.7 \\
VisualNews\_t2i & 68.4 & 75.7 & 80.3 & 74.7 & 79.5 & 45.7 & 60.2 & 63.7 & 69.1 & 74.1 \\
VisualNews\_i2t & 74.0 & 81.9 & 81.2 & 78.3 & 83.6 & 40.2 & 58.5 & 59.2 & 76.9 & 82.1 \\
MSCOCO\_t2i & 75.4 & 74.0 & 77.5 & 75.9 & 71.5 & 64.0 & 59.6 & 64.8 & 74.8 & 76.3 \\
MSCOCO\_i2t & 71.2 & 73.7 & 73.6 & 71.1 & 57.4 & 55.8 & 55.8 & 55.6 & 71.1 & 74.9 \\
NIGHTS & 67.5 & 68.2 & 67.5 & 68.4 & 67.6 & 58.8 & 64.9 & 65.2 & 66.0 & 64.6 \\
WebQA & 88.7 & 89.3 & 88.3 & 90.6 & 91.3 & 49.7 & 90.2 & 52.7 & 90.1 & 88.6 \\
FashionIQ & 35.3 & 31.8 & 16.8 & 19.6 & 37.6 & 7.2 & 8.4 & 4.0 & 16.6 & 21.4 \\
Wiki-SS-NQ & 78.9 & 80.3 & 62.1 & 67.6 & 78.4 & 50.5 & 87.2 & 60.8 & 74.6 & 83.2 \\
OVEN & 70.2 & 70.7 & 66.6 & 64.8 & 75.6 & 57.3 & 72.3 & 64.4 & 75.2 & 78.9 \\
EDIS & 87.3 & 88.9 & 85.9 & 84.2 & 96.1 & 54.9 & 81.1 & 73.8 & 84.7 & 90.4 \\
MCMR & 42.0 & 38.0 & 0.9 & 4.1 & 27.3 & 8.9 & 26.1 & 11.0 & 31.9 & 41.1 \\
MSCOCO & 66.0 & 75.3 & 75.4 & 66.2 & 31.4 & 30.3 & 31.7 & 33.3 & 46.5 & 62.0 \\
RefCOCO & 89.0 & 93.8 & 87.1 & 87.0 & 61.2 & 57.4 & 55.8 & 55.1 & 76.4 & 84.7 \\
RefCOCO-Matching & 93.1 & 93.7 & 84.4 & 86.3 & 78.7 & 83.4 & 65.8 & 97.2 & 80.3 & 92.0 \\
Visual7W-Pointing & 83.5 & 87.8 & 80.7 & 69.0 & 66.5 & 53.4 & 42.3 & 46.6 & 71.1 & 79.5 \\

\bottomrule
\end{tabular}%
}
\caption{Detailed Scores on Image Tasks.}
\label{tab:image_appendix_results}
\end{table*}

\begin{table*}[t]
\centering
\scriptsize
\setlength{\tabcolsep}{3pt}
\renewcommand{\arraystretch}{0.95}
\resizebox{\textwidth}{!}{%
\begin{tabular}{l|cccccccccc}
\toprule
\textbf{Dataset} 
& \textbf{Qwen3-VL} 
& \textbf{Qwen3-VL} 
& \textbf{VLM2Vec} 
& \textbf{VLM2Vec} 
& \textbf{GME} 
& \textbf{WAVE} 
& \textbf{Omni-Embed} 
& \textbf{LCO-Embedding} 
& \textbf{E5-Omni} 
& \textbf{E5-Omni} \\
& \textbf{-Embedding(2B)} 
& \textbf{-Embedding(8B)} 
& \textbf{-Qwen2VL (7B)} 
& \textbf{-V2.0 (2B)} 
& \textbf{(7B)} 
& \textbf{(7B)} 
& \textbf{-Nemotron (3B)} 
& \textbf{-Omni (7B)} 
& \textbf{(3B)} 
& \textbf{(7B)}  \\
\midrule
\rowcolor{blue!10}
\textbf{Avg - Video (18 tasks, Hit@1)} &55.9 &58.6 &33.8 &34.7 &38.4 &43.1 &41.4 &48.4 & 48.9 &50.8 \\

\midrule

\rowcolor{orange!15}
\textbf{V-CLS (5)} & 62.1 & 65.0 & 39.0 & 39.2 & 37.3 & 50.7 & 48.3 & 58.1 & 46.9 & 53.9 \\
\rowcolor{orange!15}
\textbf{V-QA (5)} & 61.7 & 67.6 & 30.1 & 34.7 & 50.3 & 45.9 & 47.8 & 62.4 & 60.4 & 65.9 \\
\rowcolor{orange!15}
\textbf{V-RET (5)} & 46.9 & 48.9 & 29.1 & 28.4 & 28.3 & 34.7 & 38.7 & 42.2 & 42.7 & 40.5 \\
\rowcolor{orange!15}
\textbf{V-MR (3)} & 51.2 & 49.2 & 39.2 & 37.5 & 37.5 & 39.5 & 23.5 & 19.3 & 43.4 & 37.9 \\

\midrule

K700 & 55.2 & 51.7 & 35.4 & 38.2 & 39.6 & 54.2 & 43.4 & 53.6 & 48.5 & 57.4 \\
SmthSmthV2 & 66.0 & 73.3 & 32.0 & 43.0 & 30.5 & 49.8 & 46.2 & 60.6 & 45.7 & 51.7 \\
HMDB51 & 68.8 & 75.5 & 41.9 & 40.2 & 48.0 & 54.2 & 52.1 & 52.4 & 42.6 & 56.4 \\
UCF101 & 83.7 & 85.5 & 62.1 & 60.0 & 54.4 & 74.1 & 66.6 & 75.7 & 60.5 & 71.0 \\
Breakfast & 36.7 & 38.8 & 23.6 & 14.8 & 13.9 & 21.2 & 33.3 & 48.3 & 37.2 & 32.8 \\
MVBench & 56.2 & 64.8 & 28.6 & 33.6 & 46.3 & 41.4 & 43.0 & 62.7 & 56.9 & 64.0 \\
Video-MME & 51.3 & 55.6 & 28.0 & 30.8 & 39.2 & 32.3 & 38.0 & 51.9 & 47.4 & 52.7 \\
NExTQA & 65.4 & 72.4 & 20.3 & 20.9 & 53.5 & 36.8 & 45.2 & 69.4 & 67.5 & 74.9 \\
EgoSchema & 62.6 & 64.8 & 22.2 & 35.0 & 46.4 & 47.8 & 54.8 & 49.8 & 59.0 & 57.2 \\
ActivityNetQA & 72.9 & 80.3 & 51.4 & 53.0 & 66.0 & 71.3 & 58.2 & 78.0 & 71.5 & 80.6 \\
DiDeMo & 47.9 & 44.9 & 29.6 & 30.0 & 26.3 & 29.8 & 41.7 & 43.3 & 47.7 & 42.4 \\
MSR-VTT & 48.3 & 50.1 & 34.7 & 27.8 & 31.8 & 37.3 & 40.1 & 44.7 & 45.6 & 44.8 \\
MSVD & 70.4 & 72.8 & 46.7 & 47.3 & 49.6 & 60.9 & 60.3 & 64.8 & 63.1 & 64.3 \\
VATEX & 42.5 & 45.4 & 25.4 & 26.2 & 24.7 & 34.2 & 32.8 & 36.5 & 36.5 & 32.3 \\
YouCook2 & 25.1 & 31.1 & 9.0 & 10.6 & 8.9 & 11.4 & 18.5 & 21.9 & 20.5 & 18.7 \\
QVHighlight & 74.0 & 71.3 & 57.9 & 49.7 & 59.4 & 54.5 & 20.9 & 15.1 & 49.3 & 52.9 \\
Charades-STA & 31.5 & 26.4 & 18.6 & 20.1 & 13.9 & 27.8 & 12.8 & 11.0 & 19.8 & 19.4 \\
MomentSeeker & 48.2 & 49.9 & 41.0 & 42.9 & 39.3 & 36.4 & 36.8 & 31.7 & 61.0 & 41.3 \\
\bottomrule
\end{tabular}%
}
\caption{Detailed Scores on Video Tasks.}
\label{tab:video_appendix_results}
\end{table*}

\begin{table*}[t]
\centering
\scriptsize
\setlength{\tabcolsep}{3pt}
\renewcommand{\arraystretch}{0.95}
\resizebox{\textwidth}{!}{%
\begin{tabular}{l|cccccccccc}
\toprule
\textbf{Dataset} 
& \textbf{Qwen3-VL} 
& \textbf{Qwen3-VL} 
& \textbf{VLM2Vec} 
& \textbf{VLM2Vec} 
& \textbf{GME} 
& \textbf{WAVE} 
& \textbf{Omni-Embed} 
& \textbf{LCO-Embedding} 
& \textbf{E5-Omni} 
& \textbf{E5-Omni} \\
& \textbf{-Embedding(2B)} 
& \textbf{-Embedding(8B)} 
& \textbf{-Qwen2VL (7B)} 
& \textbf{-V2.0 (2B)} 
& \textbf{(7B)} 
& \textbf{(7B)} 
& \textbf{-Nemotron (3B)} 
& \textbf{-Omni (7B)} 
& \textbf{(3B)} 
& \textbf{(7B)}  \\
\midrule
\rowcolor{blue!10}
\textbf{Avg - VisDoc (24 tasks, NDCG@5)} &70.6 &70.9 &32.6 &68.6 &75.2 &42.8 &71.1 & 68.7 &73.5 &75.4 \\

\midrule

\rowcolor{orange!15}
\textbf{ViDoRe-V1 (10)} & 81.2 & 82.1 & 20.0 & 74.4 & 89.6 & 52.3 & 84.8 & 80.1 & 84.5 & 87.3 \\
\rowcolor{orange!15}
\textbf{ViDoRe-V2 (7)} & 58.5 & 54.6 & 9.2 & 44.6 & 55.5 & 30.7 & 51.6 & 54.1 & 57.3 & 58.3 \\
\rowcolor{orange!15}
\textbf{VisRAG (6)} & 80.2 & 82.4 & 58.9 & 79.3 & 85.0 & 48.0 & 85.4 & 79.6 & 86.0 & 87.6 \\
\rowcolor{orange!15}
\textbf{VisDoc-OOD (4)} & 41.9 & 41.8 & 48.1 & 62.0 & 44.4 & 23.3 & 35.2 & 38.6 & 43.1 & 44.2 \\

\midrule

ViDoRe\_arxivqa & 77.5 & 80.0 & 28.2 & 78.9 & 87.5 & 42.0 & 83.3 & 78.1 & 83.2 & 87.9 \\
ViDoRe\_docvqa & 43.7 & 45.8 & 19.0 & 37.1 & 56.6 & 27.9 & 56.3 & 51.3 & 53.0 & 58.1 \\
ViDoRe\_infovqa & 85.2 & 85.5 & 44.8 & 82.7 & 92.2 & 64.2 & 88.4 & 85.8 & 90.2 & 92.2 \\
ViDoRe\_tabfquad & 94.9 & 94.8 & 17.0 & 87.8 & 92.7 & 68.0 & 92.3 & 92.1 & 93.4 & 93.0 \\
ViDoRe\_tatdqa & 59.5 & 59.4 & 5.7 & 44.3 & 76.6 & 19.9 & 68.9 & 58.5 & 61.9 & 70.5 \\
ViDoRe\_shiftproject & 79.4 & 81.0 & 1.6 & 61.0 & 95.6 & 43.1 & 79.7 & 71.4 & 83.2 & 84.7 \\
ViDoRe\_syntheticDocQA\_artificial\_intelligence & 95.8 & 96.4 & 18.2 & 89.1 & 99.6 & 55.6 & 95.9 & 91.9 & 98.2 & 98.9 \\
ViDoRe\_syntheticDocQA\_energy & 88.2 & 89.2 & 23.9 & 86.3 & 95.7 & 67.2 & 92.7 & 89.4 & 92.5 & 94.1 \\
ViDoRe\_syntheticDocQA\_government\_reports & 93.8 & 93.5 & 13.9 & 85.6 & 99.5 & 62.9 & 93.8 & 89.7 & 95.0 & 95.8 \\
ViDoRe\_syntheticDocQA\_healthcare\_industry & 94.1 & 95.8 & 27.6 & 91.1 & 99.5 & 71.8 & 96.3 & 92.6 & 95.0 & 98.2 \\
ViDoRe\_esg\_reports\_human\_labeled\_v2 & 57.1 & 56.6 & 7.0 & 45.8 & 62.8 & 37.2 & 60.2 & 59.3 & 63.8 & 61.3 \\
ViDoRe\_biomedical\_lectures\_v2\_multilingual & 63.5 & 61.2 & 5.2 & 44.6 & 49.8 & 27.2 & 51.2 & 57.3 & 57.0 & 60.1 \\
ViDoRe\_economics\_reports\_v2\_multilingual & 57.7 & 50.4 & 13.8 & 42.3 & 53.9 & 32.0 & 44.2 & 55.2 & 56.5 & 53.5 \\
ViDoRe\_esg\_reports\_v2\_multilingual & 55.6 & 50.2 & 11.1 & 45.7 & 55.4 & 26.3 & 50.7 & 44.8 & 52.0 & 58.4 \\

VisRAG\_ArxivQA & 79.6 & 78.7 & 52.9 & 76.7 & 87.7 & 31.6 & 83.2 & 77.0 & 84.8 & 88.1 \\
VisRAG\_ChartQA & 79.2 & 86.4 & 69.0 & 84.2 & 81.3 & 56.7 & 89.2 & 85.2 & 87.6 & 87.8 \\
VisRAG\_MP-DocVQA & 77.5 & 80.2 & 52.7 & 71.8 & 89.1 & 41.9 & 86.1 & 78.4 & 86.8 & 88.9 \\
VisRAG\_SlideVQA & 91.8 & 92.3 & 72.8 & 91.4 & 94.7 & 68.0 & 95.2 & 90.7 & 95.8 & 96.0 \\
VisRAG\_InfoVQA & 90.3 & 90.3 & 71.3 & 85.9 & 93.5 & 71.4 & 92.8 & 87.4 & 93.8 & 94.7 \\
VisRAG\_PlotQA & 62.7 & 66.2 & 34.5 & 65.9 & 63.4 & 18.4 & 65.9 & 58.7 & 67.0 & 70.1 \\
ViDoSeek-page & 22.1 & 21.9 & 77.4 & 80.3 & 23.2 & 13.6 & 1.8 & 16.1 & 22.0 & 23.4 \\
ViDoSeek-doc & 84.5 & 84.6 & 54.2 & 80.3 & 83.9 & 48.0 & 83.1 & 80.8 & 83.1 & 83.9 \\
MMLongBench-page & 16.5 & 15.9 & 36.8 & 44.7 & 16.2 & 8.3 & 5.8 & 9.2 & 15.5 & 15.6 \\
MMLongBench-doc & 44.5 & 45.0 & 24.0 & 42.9 & 54.3 & 23.4 & 50.1 & 48.2 & 51.6 & 53.9 \\

\bottomrule
\end{tabular}%
}
\caption{Detailed Scores on VisDoc Tasks.}
\label{tab:visdoc_appendix_results}
\end{table*}

\begin{table*}[t]
\centering
\scriptsize
\setlength{\tabcolsep}{4pt}
\renewcommand{\arraystretch}{0.95}
\resizebox{\textwidth}{!}{%
\begin{tabular}{l|ccccc}
\toprule
\textbf{Dataset}  
& \textbf{WAVE} 
& \textbf{Omni-Embed} 
& \textbf{LCO-Embedding} 
& \textbf{E5-Omni} 
& \textbf{E5-Omni} \\
& \textbf{(7B)} 
& \textbf{-Nemotron (3B)} 
& \textbf{-Omni (7B)} 
& \textbf{(3B)} 
& \textbf{(7B)} \\
\midrule

\rowcolor{blue!10}
\textbf{Avg - Audio (11 tasks, Hit@1)} &31.8 &36.5 &43.2 &30.8 &43.0   \\

\midrule

\rowcolor{orange!15}
\textbf{A-CLS (5)} & 52.3  & 53.7 & 64.3 & 41.4 & 63.0 \\
\rowcolor{orange!15}
\textbf{A-RET (4)} & 12.9  & 18.8 & 19.9 & 22.0 & 23.5 \\
\rowcolor{orange!15}
\textbf{A-TG (2)} & 21.5  & 29.1 & 36.8 & 21.6 & 32.2 \\

\midrule

NSynth & 28.2 & 23.0 & 53.0 & 32.9 & 44.3 \\
UrbanSound8K & 50.2 & 54.4 & 56.0 & 48.9 & 57.4 \\
ESC-50 & 83.3 & 76.5 & 66.1 & 40.0 & 68.8 \\
SpeechCommands & 43.6 & 87.0 & 84.8 & 50.4 & 86.9 \\
CREMA-D & 56.0 & 27.5 & 61.8 & 34.9 & 57.6 \\

Clotho & 14.8 & 12.3 & 14.0 & 17.5 & 19.3 \\
SoundDescs & 16.9 & 17.2 & 18.6 & 23.1 & 28.3 \\
AVE & 6.9 & 13.4 & 10.0 & 9.7 & 8.5 \\
SpeechCOCO & 7.4 & 32.2 & 37.0 & 37.7 & 37.9 \\

TUTSound & 38.4 & 53.7 & 66.9 & 39.5 & 59.3 \\
TUTSound\_hard & 4.6 & 4.4 & 6.7 & 3.8 & 5.2 \\

\bottomrule
\end{tabular}%
}
\caption{Detailed Scores on Audio Tasks.}
\label{tab:audio_appendix_results}
\end{table*}

\begin{table*}[t]
\centering
\scriptsize
\setlength{\tabcolsep}{3pt}
\renewcommand{\arraystretch}{0.95}
\resizebox{\textwidth}{!}{%
\begin{tabular}{l|ccccccccccc}
\toprule
\textbf{Dataset} 
& \textbf{Qwen3-VL} 
& \textbf{Qwen3-VL} 
& \textbf{VLM2Vec} 
& \textbf{VLM2Vec} 
& \textbf{GME} 
& \textbf{WAVE} 
& \textbf{Omni-Embed} 
& \textbf{LCO-Embedding} 
& \textbf{E5-Omni} 
& \textbf{E5-Omni} \\
& \textbf{-Embedding(2B)} 
& \textbf{-Embedding(8B)} 
& \textbf{-Qwen2VL (7B)} 
& \textbf{-V2.0 (2B)} 
& \textbf{(7B)} 
& \textbf{(7B)} 
& \textbf{-Nemotron (3B)} 
& \textbf{-Omni (7B)} 
& \textbf{(3B)} 
& \textbf{(7B)}  \\
\midrule
\rowcolor{blue!10}
\textbf{Avg - Text (53 tasks, NDCG@5)} & 40.1 & 43.3 & 22.6 & 25.1 & 38.2 & 14.1 & 39.1 & 33.0 & 27.2 & 27.4 \\

\midrule

\rowcolor{orange!15}
\textbf{T-RR (20)} & 16.6 & 18.2 & 7.2 & 7.8 & 12.5 & 5.9 & 17.2 & 13.6 & 11.5 & 11.1 \\
\rowcolor{orange!15}
\textbf{T-IF (9)} & 46.0 & 49.5 & 30.2 & 32.8 & 58.6 & 34.0 & 42.3 & 55.4 & 47.4 & 44.8 \\
\rowcolor{orange!15}
\textbf{T-LR (6)} & 53.1 & 58.0 & 5.9 & 11.5 & 17.8 & 2.6 & 50.0 & 12.6 & 8.3 & 9.7 \\
\rowcolor{orange!15}
\textbf{T-MR (5)} & 61.2 & 61.2 & 40.9 & 50.3 & 59.0 & 22.9 & 69.7 & 61.4 & 60.7 & 56.9 \\
\rowcolor{orange!15}
\textbf{T-GR (13)} & 58.0 & 61.2 & 41.7 & 41.2 & 62.5 & 18.6 & 56.9 & 52.1 & 34.4 & 35.9 \\

\midrule

BRIGHT\_aops & 2.4 & 1.1 & 2.2 & 1.2 & 5.0 & 5.3 & 3.3 & 8.2 & 3.7 & 3.9 \\
BRIGHT\_biology & 14.7 & 17.1 & 9.4 & 12.3 & 3.0 & 2.7 & 13.6 & 11.1 & 2.0 & 1.6 \\
BRIGHT\_earth\_science & 26.2 & 28.9 & 0.7 & 11.4 & 12.8 & 3.9 & 26.4 & 13.6 & 7.7 & 6.1 \\
BRIGHT\_economics & 15.9 & 16.7 & 2.8 & 3.9 & 7.0 & 2.1 & 15.4 & 10.3 & 4.5 & 3.8 \\
BRIGHT\_leetcode & 13.7 & 17.3 & 4.7 & 8.5 & 17.3 & 12.0 & 18.5 & 5.6 & 17.9 & 20.3 \\
BRIGHT\_pony & 2.5 & 9.1 & 9.5 & 1.5 & 2.1 & 0.5 & 2.6 & 0.5 & 2.4 & 1.3 \\
BRIGHT\_psychology & 18.6 & 12.0 & 3.0 & 4.4 & 11.0 & 2.8 & 18.1 & 8.5 & 5.6 & 4.2 \\
BRIGHT\_robotics & 11.3 & 12.5 & 1.5 & 4.8 & 5.7 & 0.7 & 12.1 & 2.8 & 7.9 & 7.0 \\
BRIGHT\_stackoverflow & 14.5 & 12.0 & 0.0 & 2.1 & 6.1 & 0.9 & 9.6 & 1.6 & 6.8 & 6.0 \\
BRIGHT\_sustainable\_living & 9.7 & 14.2 & 3.6 & 2.6 & 9.7 & 5.1 & 6.6 & 10.9 & 4.7 & 4.9 \\
BRIGHT\_theoremqa\_questions & 10.6 & 12.8 & 15.1 & 9.4 & 12.8 & 7.7 & 13.9 & 9.9 & 12.3 & 16.9 \\
BRIGHT\_theoremqa\_theorems & 12.8 & 21.2 & 3.2 & 4.7 & 17.0 & 1.9 & 23.0 & 8.6 & 11.2 & 11.6 \\

FollowIR\_core17-instructions & 45.3 & 44.0 & 26.1 & 27.5 & 50.0 & 29.1 & 20.7 & 52.2 & 34.8 & 32.4 \\
FollowIR\_news21-instructions & 47.7 & 46.7 & 17.6 & 26.4 & 44.4 & 25.3 & 43.2 & 46.3 & 41.6 & 42.3 \\
FollowIR\_robust04-instructions & 50.8 & 51.1 & 26.7 & 29.4 & 40.1 & 12.1 & 26.2 & 44.5 & 36.5 & 29.3 \\

InfoSearch\_Audience-v1 & 31.3 & 34.5 & 25.0 & 17.2 & 61.8 & 31.9 & 27.4 & 48.6 & 29.1 & 34.2 \\
InfoSearch\_Clarity-v1 & 62.8 & 55.4 & 34.5 & 43.6 & 57.9 & 59.2 & 34.5 & 50.5 & 62.3 & 43.8 \\
InfoSearch\_Format-v1 & 38.2 & 46.2 & 10.2 & 20.4 & 47.5 & 7.2 & 52.5 & 47.3 & 49.9 & 49.2 \\
InfoSearch\_Language-v1 & 56.5 & 67.6 & 45.3 & 49.2 & 77.5 & 60.8 & 71.3 & 68.9 & 67.9 & 66.5 \\
InfoSearch\_Length-v1 & 44.2 & 55.9 & 40.0 & 43.9 & 75.4 & 38.0 & 56.9 & 73.3 & 56.4 & 58.6 \\
InfoSearch\_Source-v1 & 36.8 & 44.5 & 46.1 & 37.6 & 72.5 & 42.1 & 47.7 & 67.0 & 47.8 & 46.9 \\

LongEmbed\_2wikimqa & 83.8 & 89.0 & 6.0 & 13.7 & 14.7 & 1.9 & 71.1 & 2.5 & 1.4 & 1.2 \\
LongEmbed\_narrativeqa & 52.9 & 60.9 & 1.3 & 3.6 & 6.8 & 1.2 & 32.7 & 1.8 & 1.2 & 1.1 \\
LongEmbed\_needle & 15.5 & 21.2 & 3.2 & 7.6 & 21.2 & 5.0 & 24.3 & 17.1 & 14.9 & 21.7 \\
LongEmbed\_passkey & 32.8 & 38.2 & 12.7 & 11.3 & 26.7 & 1.1 & 35.9 & 31.8 & 27.7 & 31.4 \\
LongEmbed\_qmsum & 37.5 & 42.3 & 4.7 & 11.9 & 9.8 & 2.4 & 40.2 & 8.4 & 1.5 & 1.5 \\
LongEmbed\_summ\_screen\_fd & 95.9 & 96.3 & 7.5 & 20.9 & 27.4 & 3.8 & 96.0 & 13.8 & 3.1 & 1.3 \\

MultiConIR\_Books & 62.1 & 60.9 & 35.0 & 47.1 & 56.8 & 27.7 & 64.0 & 56.1 & 52.1 & 52.5 \\
MultiConIR\_Legal Document & 57.2 & 57.7 & 38.0 & 52.4 & 56.7 & 8.8 & 68.8 & 59.8 & 57.7 & 54.7 \\
MultiConIR\_Medical Case & 60.0 & 61.4 & 35.8 & 43.9 & 52.8 & 17.5 & 65.7 & 59.4 & 57.7 & 51.6 \\
MultiConIR\_Movies & 64.9 & 62.9 & 42.0 & 48.7 & 64.6 & 13.0 & 74.7 & 65.2 & 66.3 & 61.9 \\
MultiConIR\_People & 61.8 & 62.9 & 53.4 & 59.2 & 64.1 & 47.3 & 75.1 & 66.2 & 69.8 & 64.0 \\

NanoBEIR\_NanoArguAna & 49.4 & 45.0 & 32.4 & 36.1 & 54.3 & 21.6 & 42.8 & 25.7 & 53.3 & 49.9 \\
NanoBEIR\_NanoClimateFEVER & 29.5 & 33.5 & 14.2 & 14.4 & 31.9 & 3.4 & 23.6 & 25.0 & 2.2 & 1.3 \\
NanoBEIR\_NanoDBPedia & 60.6 & 61.1 & 51.7 & 48.4 & 60.5 & 23.6 & 53.8 & 65.8 & 38.2 & 36.2 \\
NanoBEIR\_NanoFEVER & 89.2 & 89.8 & 55.9 & 52.3 & 87.1 & 10.5 & 75.4 & 80.0 & 18.6 & 19.0 \\
NanoBEIR\_NanoFiQA2018 & 43.7 & 53.2 & 20.1 & 21.0 & 58.8 & 4.8 & 54.4 & 43.2 & 15.4 & 24.0 \\
NanoBEIR\_NanoHotpotQA & 73.8 & 77.4 & 53.6 & 50.0 & 84.5 & 15.3 & 92.0 & 62.2 & 73.0 & 56.9 \\
NanoBEIR\_NanoMSMARCO & 53.6 & 64.9 & 28.1 & 38.6 & 65.5 & 19.0 & 53.4 & 45.7 & 47.0 & 44.4 \\
NanoBEIR\_NanoNFCorpus & 39.5 & 39.9 & 31.8 & 29.5 & 39.9 & 11.7 & 31.8 & 34.8 & 10.8 & 13.2 \\
NanoBEIR\_NanoNQ & 56.0 & 66.6 & 34.5 & 28.2 & 65.9 & 12.5 & 64.5 & 58.5 & 28.8 & 23.9 \\
NanoBEIR\_NanoQuoraRetrieval & 94.9 & 96.0 & 93.4 & 92.8 & 95.2 & 87.1 & 91.5 & 90.5 & 92.9 & 87.5 \\
NanoBEIR\_NanoSCIDOCS & 35.7 & 37.5 & 30.1 & 24.3 & 44.1 & 12.2 & 32.7 & 33.0 & 7.2 & 33.1 \\
NanoBEIR\_NanoSciFact & 77.5 & 78.5 & 61.7 & 58.4 & 73.4 & 14.4 & 73.8 & 66.9 & 33.5 & 43.4 \\
NanoBEIR\_NanoTouche2020 & 50.1 & 52.0 & 34.2 & 41.6 & 52.0 & 5.6 & 50.3 & 46.3 & 25.7 & 33.6 \\

R2MED\_Bioinformatics & 35.8 & 42.2 & 15.1 & 19.4 & 33.2 & 2.5 & 36.9 & 14.0 & 17.5 & 33.7 \\
R2MED\_Biology & 14.7 & 17.2 & 9.3 & 12.3 & 3.2 & 2.7 & 13.9 & 11.0 & 2.0 & 1.6 \\
R2MED\_IIYi-Clinical & 21.3 & 26.3 & 2.3 & 8.5 & 18.5 & 0.1 & 15.7 & 4.2 & 15.5 & 21.9 \\
R2MED\_MedQA-Diag & 33.6 & 34.4 & 16.4 & 21.8 & 28.9 & 17.8 & 6.9 & 6.7 & 8.4 & 6.8 \\
R2MED\_MedXpertQA-Exam & 11.5 & 17.4 & 6.0 & 5.6 & 5.3 & 0.5 & 5.3 & 6.5 & 5.3 & 6.9 \\
R2MED\_Medical-Sciences & 10.1 & 13.1 & 3.5 & 4.7 & 9.3 & 0.8 & 27.3 & 33.1 & 25.8 & 20.0 \\
R2MED\_PMC-Clinical & 21.4 & 37.1 & 13.5 & 14.0 & 32.2 & 1.0 & 21.7 & 8.7 & 28.8 & 25.7 \\
R2MED\_PMC-Treatment & 30.4 & 37.4 & 21.4 & 28.3 & 42.0 & 1.1 & 14.2 & 13.9 & 27.6 & 34.9 \\

\bottomrule
\end{tabular}%
}
\caption{Detailed Scores on Text Tasks.}
\label{tab:text_appendix_results}
\end{table*}

\begin{table*}[t]
\centering
\scriptsize
\setlength{\tabcolsep}{3pt}
\renewcommand{\arraystretch}{0.95}
\resizebox{\textwidth}{!}{%
\begin{tabular}{l|cccccccccc}
\toprule
\textbf{Dataset} 
& \textbf{Qwen3-VL} 
& \textbf{Qwen3-VL} 
& \textbf{VLM2Vec} 
& \textbf{VLM2Vec} 
& \textbf{GME} 
& \textbf{WAVE} 
& \textbf{Omni-Embed} 
& \textbf{LCO-Embedding} 
& \textbf{E5-Omni} 
& \textbf{E5-Omni} \\
& \textbf{-Embedding(2B)} 
& \textbf{-Embedding(8B)} 
& \textbf{-Qwen2VL (7B)} 
& \textbf{-V2.0 (2B)} 
& \textbf{(7B)} 
& \textbf{(7B)} 
& \textbf{-Nemotron (3B)} 
& \textbf{-Omni (7B)} 
& \textbf{(3B)} 
& \textbf{(7B)} \\
\midrule
\rowcolor{blue!10}
\textbf{Avg - Agent (47 tasks, Hit@1)} &39.3 &38.4 &19.7 &28.7 &35.6 &11.3 &36.5 & 27.8 &36.9 &36.7 \\

\midrule

\rowcolor{orange!15}
\textbf{Tool (35)} & 42.6 & 41.3 & 19.8 & 27.6 & 39.0 & 11.9 & 38.1 & 29.0 & 37.7 & 37.4 \\
\rowcolor{orange!15}
\textbf{GUI (8)} & 30.4 & 33.5 & 21.4 & 36.2 & 30.0 & 11.8 & 32.0 & 25.0 & 35.6 & 38.0 \\
\rowcolor{orange!15}
\textbf{Memory (4)} & 28.4 & 22.8 & 15.9 & 23.3 & 17.1 & 5.7 & 32.2 & 23.0 & 32.3 & 27.3 \\

\midrule

Tool-craft-math-algebra & 84.6 & 83.9 & 57.1 & 65.0 & 84.6 & 17.5 & 85.0 & 63.9 & 75.7 & 81.4 \\
Tool-craft-tabmwp & 36.8 & 34.5 & 10.9 & 14.4 & 27.6 & 4.0 & 29.3 & 25.3 & 19.0 & 29.9 \\
Tool-craft-vqa & 52.0 & 54.5 & 23.0 & 43.5 & 60.5 & 12.5 & 65.5 & 34.5 & 47.0 & 50.5 \\
Tool-gorilla-huggingface & 26.4 & 28.4 & 18.6 & 20.2 & 24.6 & 6.4 & 26.6 & 20.2 & 25.8 & 19.2 \\
Tool-gorilla-pytorch & 11.6 & 9.3 & 0.0 & 4.7 & 14.0 & 2.3 & 11.6 & 16.3 & 20.9 & 14.0 \\
Tool-gorilla-tensor & 16.4 & 16.4 & 1.8 & 7.3 & 21.8 & 1.8 & 20.0 & 5.5 & 16.4 & 12.7 \\
Tool-toolink & 67.8 & 67.6 & 36.4 & 40.4 & 66.4 & 25.0 & 67.0 & 65.6 & 68.4 & 70.6 \\

Tool-apibank & 62.4 & 58.4 & 1.0 & 38.6 & 52.5 & 0.0 & 45.5 & 7.9 & 53.5 & 47.5 \\
Tool-apigen & 55.4 & 55.5 & 49.3 & 46.5 & 54.0 & 23.8 & 64.0 & 50.2 & 59.7 & 61.2 \\
Tool-mnms & 36.4 & 30.3 & 27.3 & 30.3 & 30.3 & 21.2 & 42.4 & 27.3 & 54.5 & 51.5 \\
Tool-reversechain & 61.0 & 57.5 & 22.5 & 51.5 & 54.5 & 0.5 & 50.0 & 18.5 & 64.0 & 56.5 \\
Tool-rotbench & 3.3 & 4.2 & 1.1 & 1.8 & 7.3 & 1.1 & 6.9 & 2.5 & 5.1 & 4.2 \\
Tool-t-eval-dialog & 24.0 & 20.0 & 26.0 & 22.0 & 34.0 & 0.0 & 14.0 & 20.0 & 16.0 & 24.0 \\
Tool-t-eval-step & 16.0 & 12.0 & 38.0 & 10.0 & 30.0 & 8.0 & 12.0 & 22.0 & 26.0 & 12.0 \\
Tool-taskbench-daily & 65.0 & 62.5 & 45.0 & 52.5 & 72.5 & 2.5 & 67.5 & 62.5 & 90.0 & 75.0 \\
Tool-toolace & 65.7 & 62.3 & 33.3 & 56.4 & 69.0 & 24.3 & 67.0 & 51.7 & 67.1 & 71.4 \\
Tool-toolbench & 40.6 & 36.0 & 6.0 & 24.8 & 17.4 & 4.6 & 13.3 & 16.9 & 29.2 & 28.9 \\
Tool-toolemu & 31.6 & 26.3 & 0.0 & 10.5 & 13.2 & 0.0 & 34.2 & 7.9 & 10.5 & 5.3 \\
Tool-tooleyes & 19.0 & 20.0 & 4.2 & 15.8 & 20.0 & 0.0 & 10.5 & 11.6 & 24.2 & 12.6 \\
Tool-toollens & 8.0 & 7.0 & 1.3 & 8.9 & 1.3 & 1.6 & 1.0 & 2.9 & 5.1 & 10.2 \\
Tool-ultratool & 50.6 & 41.2 & 0.4 & 4.6 & 40.4 & 5.2 & 43.6 & 41.2 & 44.0 & 43.8 \\
Tool-autotools-food & 50.0 & 50.0 & 9.1 & 4.6 & 18.2 & 0.0 & 13.6 & 4.5 & 9.1 & 18.2 \\
Tool-autotools-music & 12.5 & 37.5 & 0.0 & 0.0 & 0.0 & 0.0 & 3.1 & 0.0 & 0.0 & 0.0 \\
Tool-autotools-weather & 18.2 & 9.1 & 0.0 & 0.0 & 0.0 & 0.0 & 0.0 & 0.0 & 0.0 & 9.1 \\
Tool-restgpt-spotify & 60.0 & 67.5 & 27.5 & 12.5 & 20.0 & 0.0 & 25.0 & 0.0 & 20.0 & 30.0 \\
Tool-restgpt-tmdb & 37.0 & 38.9 & 0.0 & 0.0 & 11.1 & 0.0 & 16.7 & 9.3 & 9.3 & 13.0 \\

Tool-appbench & 71.9 & 84.4 & 65.6 & 65.6 & 81.2 & 71.9 & 87.5 & 59.4 & 87.5 & 84.4 \\
Tool-gpt4tools & 84.4 & 68.8 & 28.1 & 46.9 & 81.2 & 50.0 & 75.0 & 81.2 & 37.5 & 50.0 \\
Tool-gta & 14.3 & 21.4 & 14.3 & 0.0 & 35.7 & 0.0 & 42.9 & 28.6 & 28.6 & 14.3 \\
Tool-taskbench-huggingface & 60.9 & 39.1 & 4.3 & 56.5 & 56.5 & 8.7 & 52.2 & 52.2 & 47.8 & 56.5 \\
Tool-taskbench-multimedia & 70.0 & 62.5 & 32.5 & 65.0 & 87.5 & 42.5 & 77.5 & 60.0 & 92.5 & 82.5 \\
Tool-metatool & 52.5 & 51.5 & 32.0 & 44.0 & 54.0 & 26.5 & 49.0 & 48.5 & 46.0 & 51.0 \\
Tool-tool-be-honest & 51.1 & 49.1 & 47.1 & 46.0 & 55.4 & 23.1 & 54.3 & 35.4 & 54.0 & 52.0 \\
Tool-toolalpaca & 62.8 & 67.0 & 26.6 & 51.1 & 59.6 & 27.7 & 50.0 & 60.6 & 58.5 & 60.6 \\
Tool-toolbench-sam & 11.7 & 11.7 & 1.0 & 4.1 & 7.6 & 3.0 & 8.1 & 2.5 & 5.6 & 6.6 \\

GAE-guiact-q2s & 19.8 & 29.5 & 7.7 & 25.9 & 31.3 & 5.5 & 26.1 & 24.1 & 29.9 & 32.3 \\
GAE-guiact-q2t & 48.4 & 46.3 & 42.6 & 52.6 & 31.3 & 7.4 & 39.8 & 42.8 & 50.9 & 55.5 \\
GAE-guiact-s2s & 25.5 & 30.2 & 33.2 & 35.8 & 32.7 & 17.2 & 35.9 & 17.9 & 36.5 & 36.4 \\
GAE-guiact-t2s & 25.0 & 24.9 & 8.6 & 24.1 & 17.1 & 5.9 & 20.7 & 13.8 & 21.9 & 23.0 \\
GAE-mind2web-q2s & 27.1 & 37.7 & 12.2 & 43.0 & 42.1 & 10.9 & 41.7 & 33.0 & 43.0 & 44.9 \\
GAE-mind2web-q2t & 44.3 & 37.2 & 32.4 & 47.8 & 32.4 & 11.9 & 33.6 & 25.3 & 40.3 & 50.2 \\
GAE-mind2web-s2s & 30.2 & 35.2 & 32.4 & 37.1 & 37.1 & 28.0 & 37.1 & 25.9 & 38.3 & 37.7 \\
GAE-mind2web-t2s & 23.1 & 27.4 & 1.9 & 23.4 & 15.9 & 7.8 & 21.5 & 17.1 & 23.7 & 23.7 \\
KnowMeBench & 41.5 & 35.3 & 22.3 & 32.6 & 19.0 & 11.6 & 44.4 & 35.8 & 45.1 & 38.4 \\
REALTALK & 35.4 & 30.5 & 11.8 & 18.4 & 19.6 & 6.6 & 32.0 & 28.1 & 27.8 & 27.5 \\
PeerQA & 17.7 & 16.2 & 11.0 & 21.3 & 14.7 & 3.7 & 18.4 & 15.4 & 22.8 & 19.9 \\
DeepPlanning & 19.2 & 9.2 & 18.3 & 20.8 & 15.0 & 0.8 & 34.2 & 12.5 & 33.3 & 23.3 \\

\bottomrule
\end{tabular}%
}
\caption{Detailed Scores on Agent Tasks.}
\label{tab:agent_appendix_results}
\end{table*}










\subsection{Visualization and Analysis}
In this section, we provide additional visualizations to complement the analysis in the main text. These figures illustrate both the global structure of the embedding space and the behavior of instruction-induced shifts across different models.

\begin{figure*}[t]
    \centering
    \begin{subfigure}{0.48\textwidth}
        \centering
        \includegraphics[width=\linewidth]{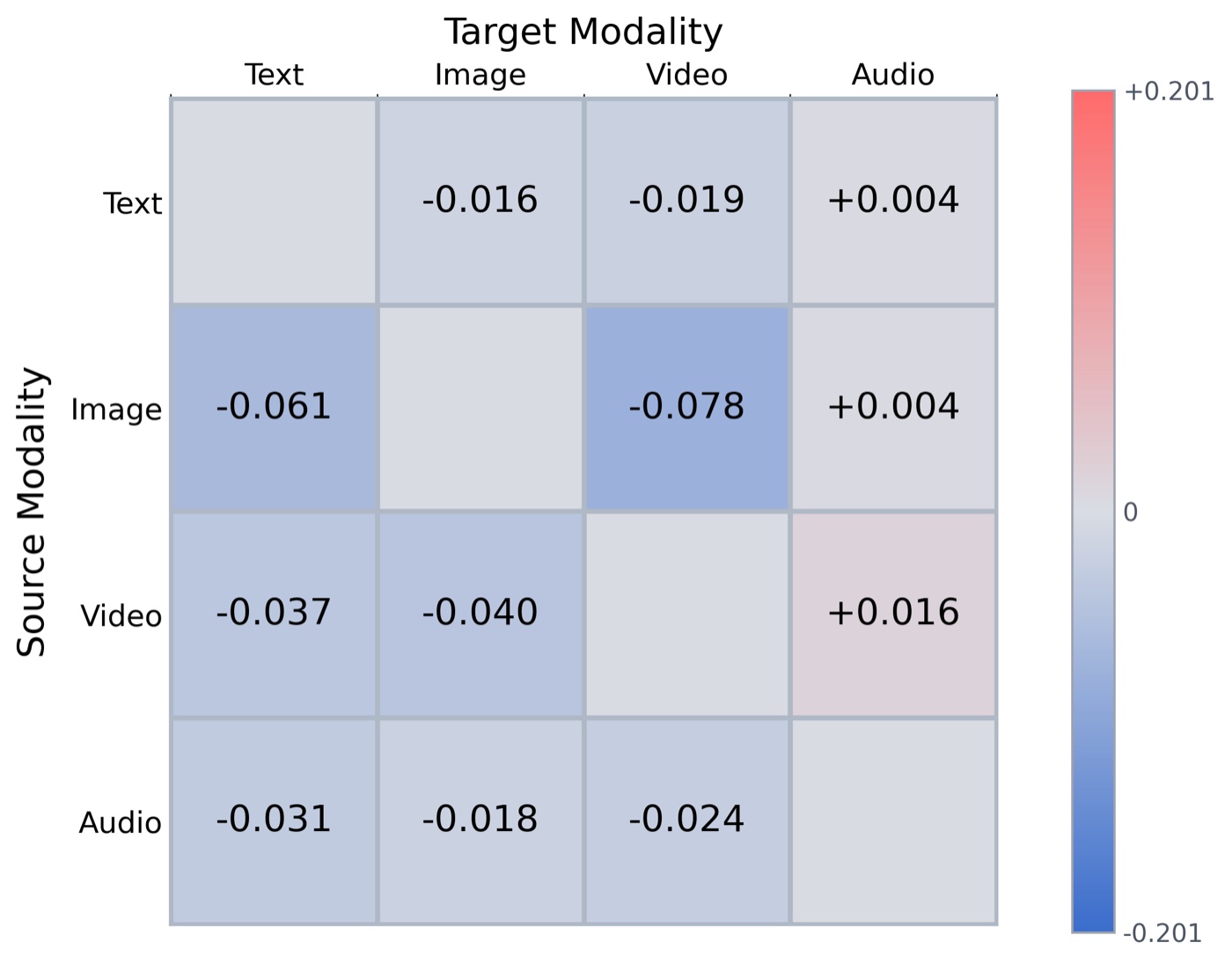}
        \caption{\texttt{WAVE}.}
        \label{fig:wave-heatmap}
    \end{subfigure}
    \hfill
    \begin{subfigure}{0.48\textwidth}
        \centering
        \includegraphics[width=\linewidth]{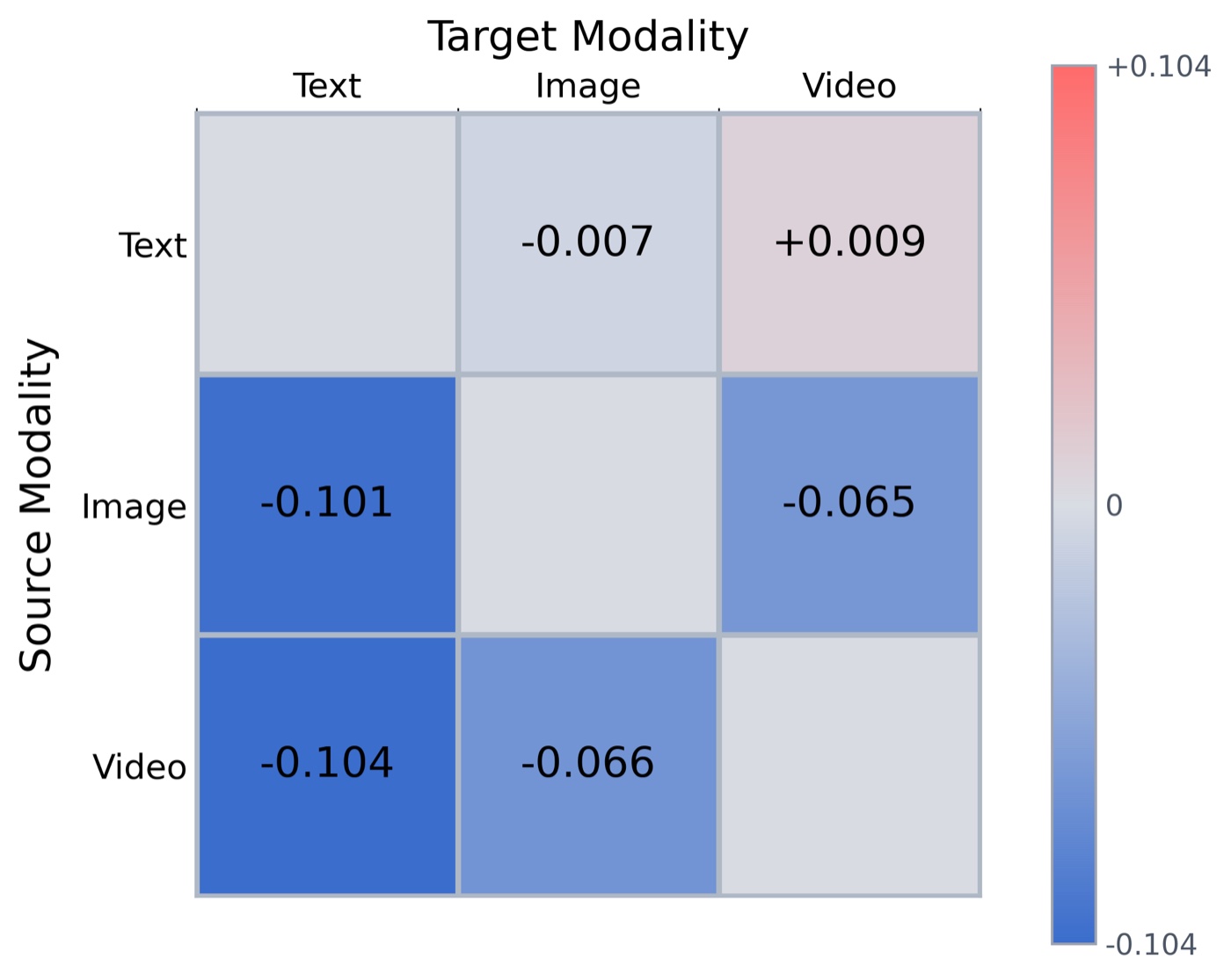}
        \caption{\texttt{Qwen3-VL-Embedding-8B}.}
        \label{fig:qwen3-heatmap}
    \end{subfigure}
\caption{
\textbf{Instruction-induced changes in query--target distance across models.}
Each cell shows the change in cosine distance to the target modality after instruction augmentation, computed relative to the raw query. Negative values indicate that the instruction-augmented query moves closer to the target modality, while positive values indicate increased distance.
}
\label{fig:instruction_target_shift_wave_}
\end{figure*}

\paragraph{Instruction-induced changes in query--target distance.}
Figure~\ref{fig:instruction_target_shift_wave_} presents the change in cosine distance between queries and target modalities after instruction augmentation for \texttt{WAVE} and \texttt{Qwen3-VL-Embedding-8B}. Compared to \texttt{omni-embed-nemotron-3b} in the main text, \texttt{WAVE} exhibits consistently small changes across all directions, indicating that instruction augmentation has limited influence on the query representation. In contrast, \texttt{Qwen3-VL-Embedding-8B} shows moderate variations, but the improvements are sparse and inconsistent. In most directions, the distance to the target modality either remains unchanged or increases, suggesting that instruction signals do not reliably improve target alignment. These observations are consistent with the findings in Section~\ref{sec:analysis}, where we show that both insufficient sensitivity and misaligned shifts contribute to retrieval failures.

\begin{figure*}[t]
    \centering
    \begin{subfigure}{0.32\textwidth}
        \centering
        \includegraphics[width=\linewidth]{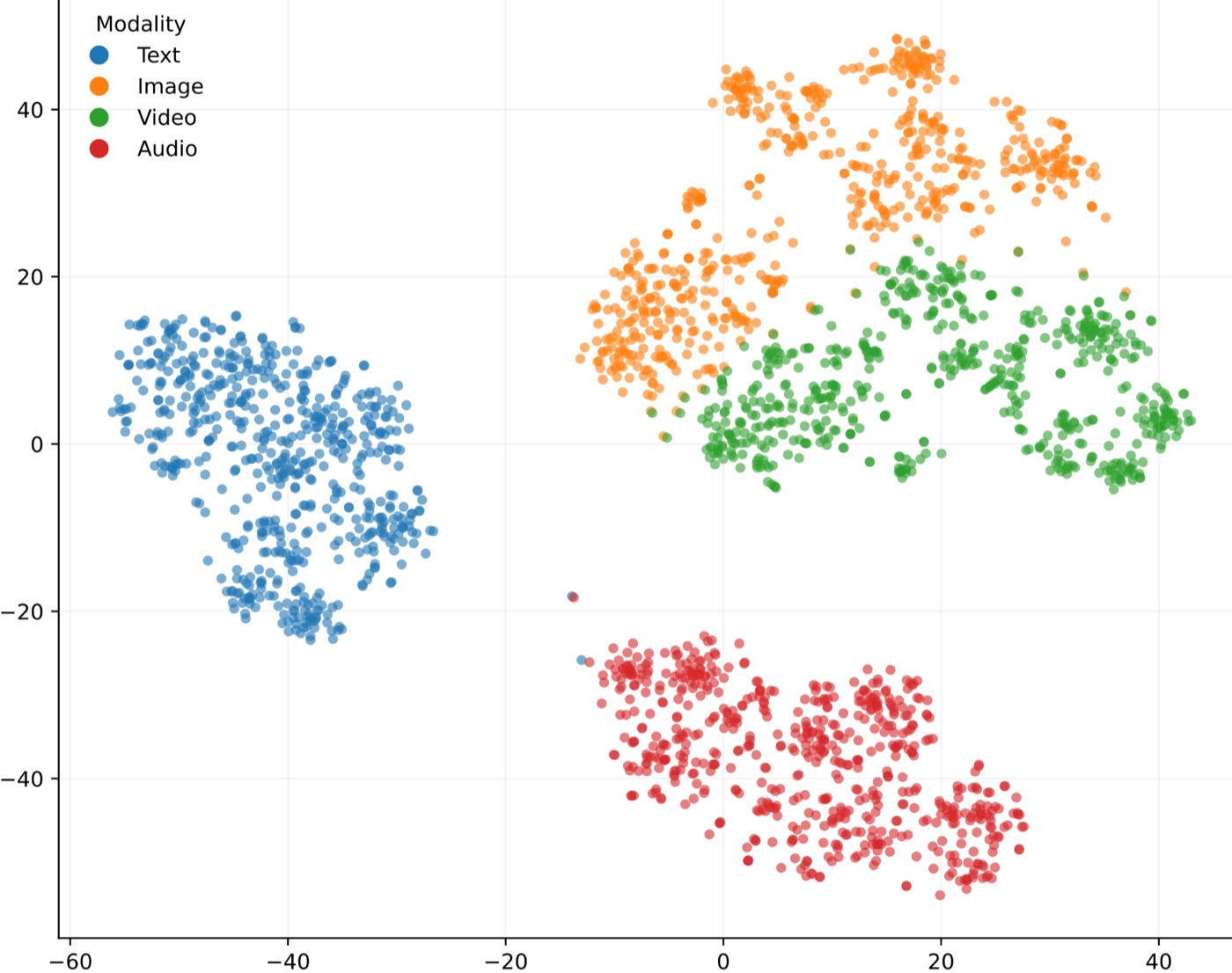}
        \caption{\texttt{omni-embed-nemotron-3b}.}
        \label{fig:embedding_space_nemotron}
    \end{subfigure}
    \hfill
    \begin{subfigure}{0.32\textwidth}
        \centering
        \includegraphics[width=\linewidth]{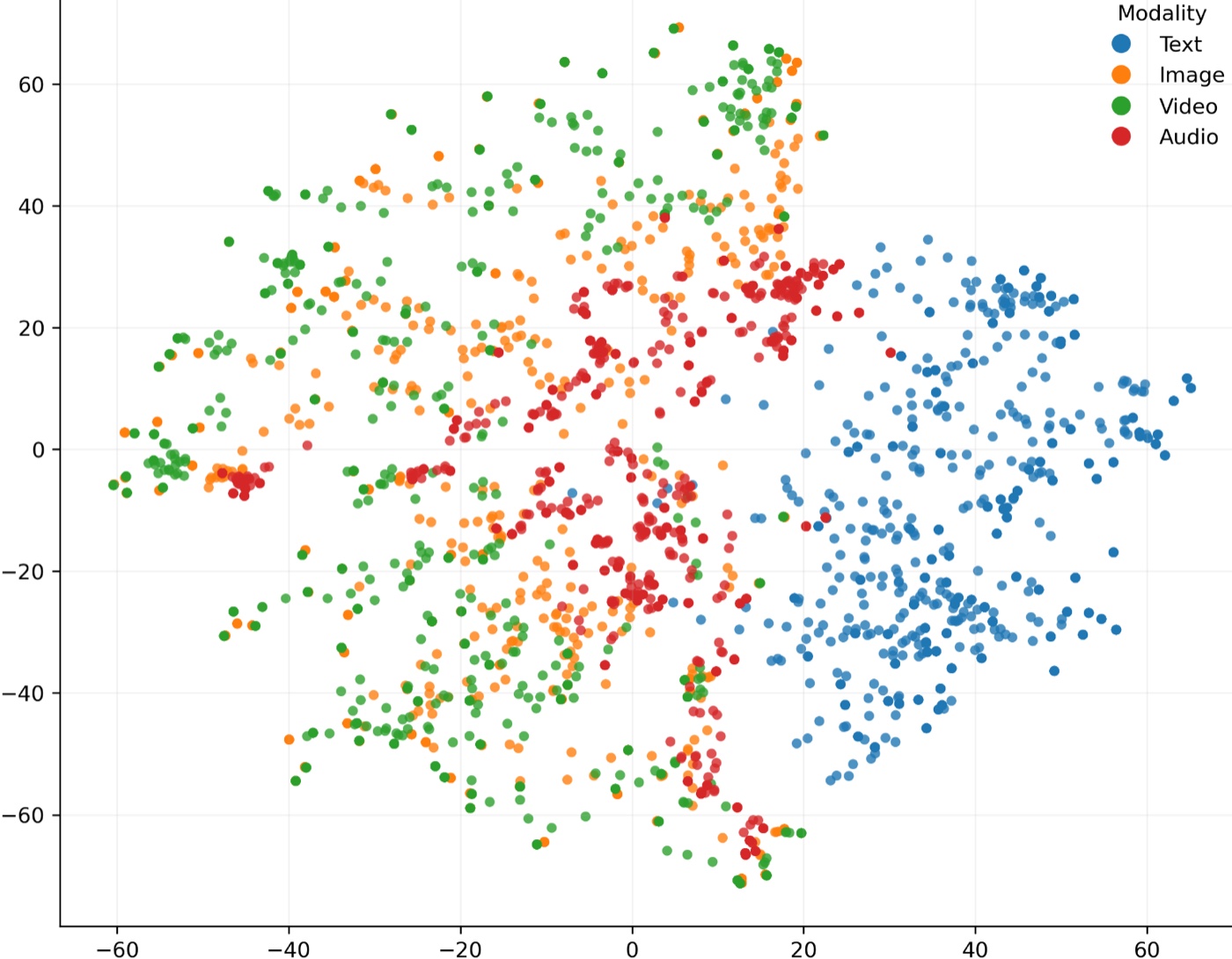}
        \caption{\texttt{WAVE}.}
        \label{fig:embedding_space_wave}
    \end{subfigure}
    \hfill
    \begin{subfigure}{0.32\textwidth}
        \centering
        \includegraphics[width=\linewidth]{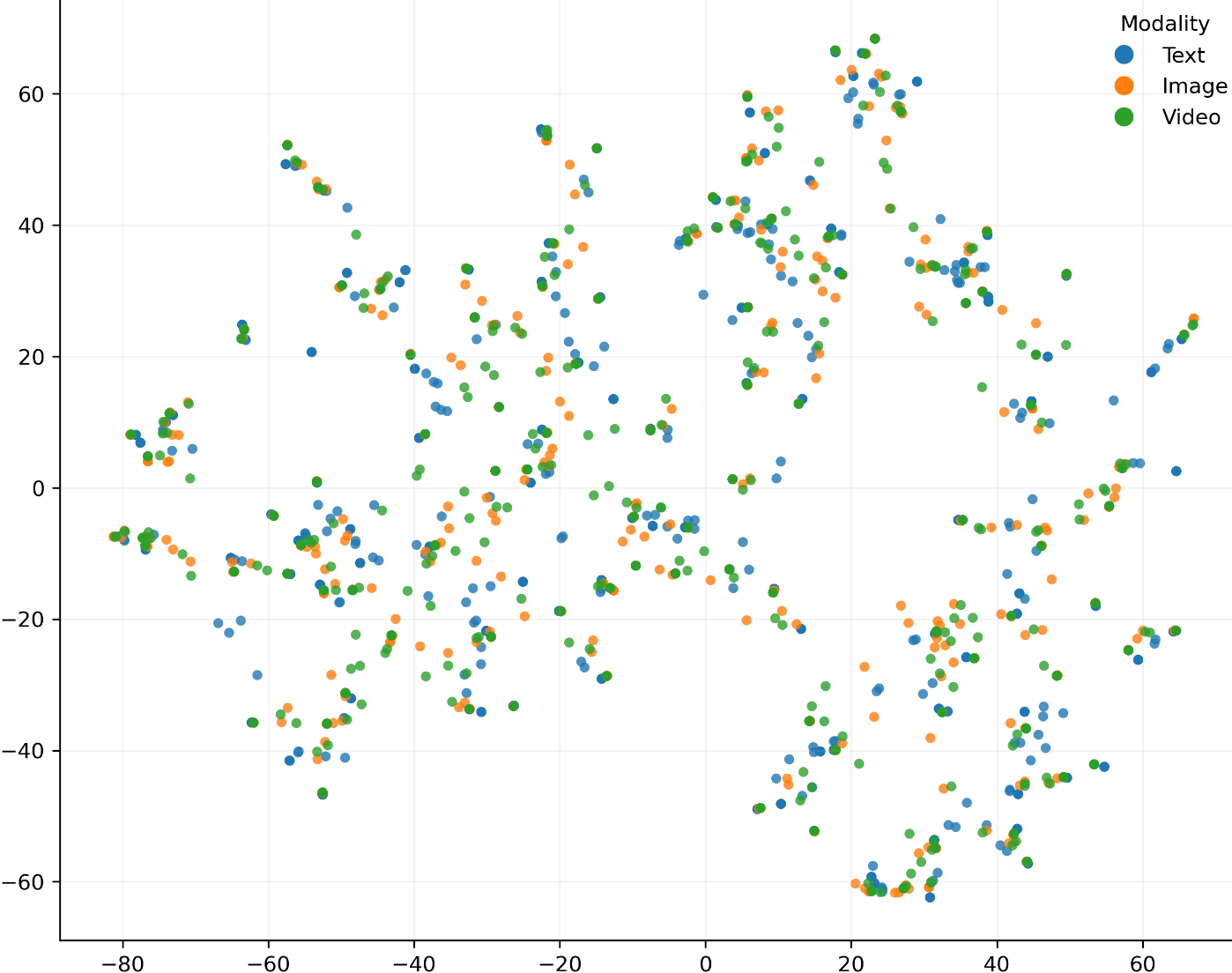}
        \caption{\texttt{Qwen3-VL-Embedding-8B}.}
        \label{fig:embedding_space_qwen3}
    \end{subfigure}
    \caption{
    \textbf{Embedding space geometry across models.}
    t-SNE visualizations of embeddings from different modalities for three representative models.
    }
    \label{fig:embedding_space_models}
\end{figure*}

\paragraph{Embedding space geometry across models.}
Figure~\ref{fig:embedding_space_models} visualizes the embedding distributions of different modalities using t-SNE. For \texttt{omni-embed-nemotron-3b}, embeddings form relatively well-separated clusters corresponding to each modality, indicating a structured representation space. \texttt{WAVE} exhibits a more entangled distribution, particularly between audio and video modalities, reflecting its focus on audio--visual representation learning. In contrast, \texttt{Qwen3-VL-Embedding-8B} shows less clearly separated clusters, with significant overlap among modalities. This difference in geometry suggests that models vary in how modality information is encoded, which in turn influences their behavior under modality-constrained retrieval.

\begin{figure*}[t]
    \centering
    \begin{subfigure}{0.48\textwidth}
        \centering
        \includegraphics[width=\linewidth]{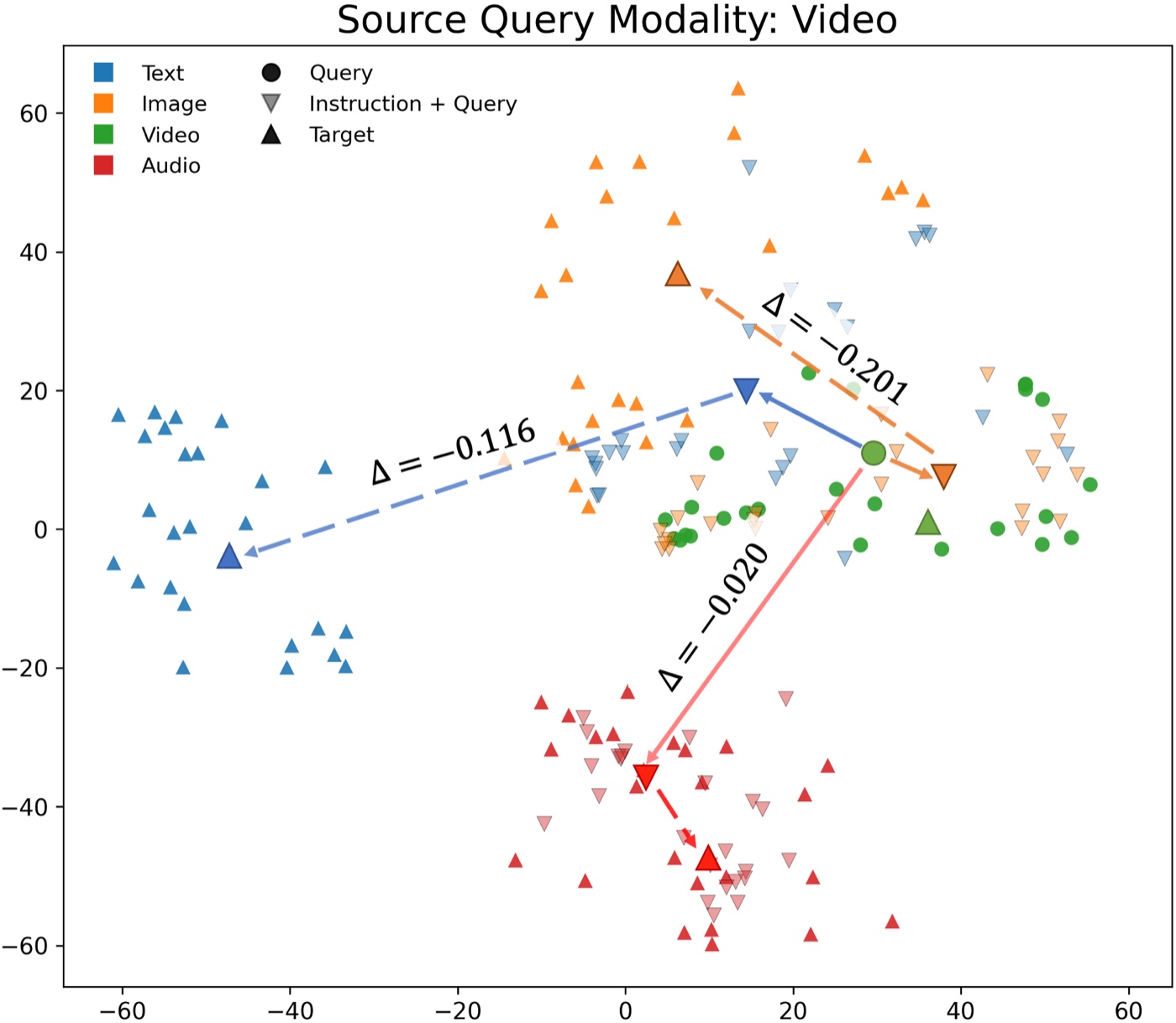}
        \caption{Query modality: video.}
        \label{fig:nemotron_video}
    \end{subfigure}
    \hfill
    \begin{subfigure}{0.48\textwidth}
        \centering
        \includegraphics[width=\linewidth]{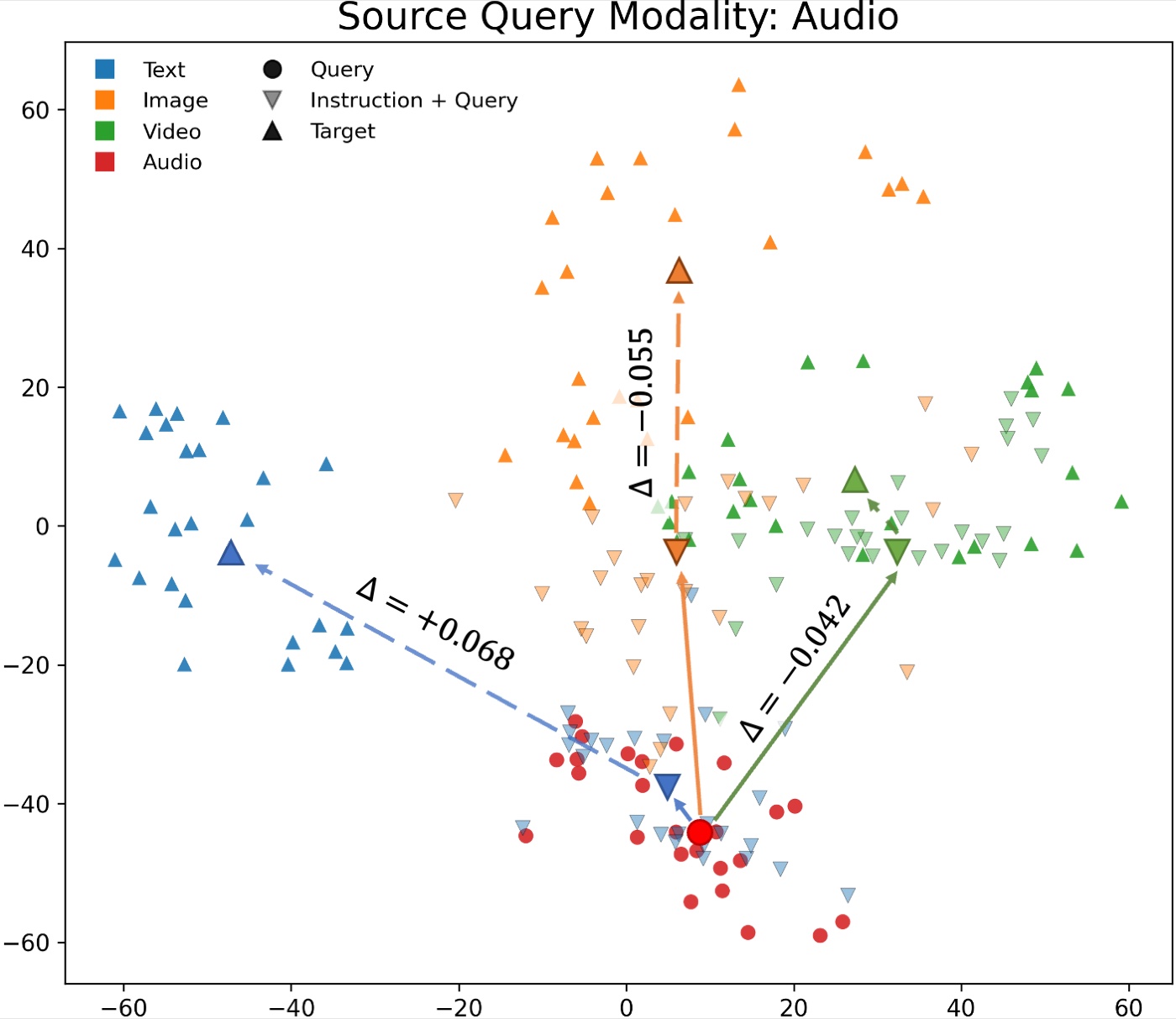}
        \caption{Query modality: audio.}
        \label{fig:nemotron_audio}
    \end{subfigure}
    \caption{
    Instruction-induced shifts in the embedding space for different source query modalities (\texttt{omni-embed-nemotron-3b}).
    Across different source modalities, instruction augmentation perturbs the query representations but does not consistently move them toward the target clusters.
    }
    \label{fig:nemotron_instruction_shift_tva}
\end{figure*}

\paragraph{Instruction-induced shifts across query modalities.}
Figure~\ref{fig:nemotron_instruction_shift_tva} further illustrates the effect of instruction augmentation on query embeddings for \texttt{omni-embed-nemotron-3b}, using video and audio queries as examples. Across different source modalities, instruction augmentation consistently perturbs the query representations. However, these shifts are not consistently directed toward the intended target modality. Instead, the updated query embeddings may move toward neighboring modality clusters or follow directions that are unrelated to the target. This observation provides additional qualitative evidence that, although the model is sensitive to instruction signals, the induced shifts do not reliably support target-oriented alignment.
\label{sec:visualiztion}

\end{document}